\documentclass[superscriptaddress, twocolumn, amsmath, amssymb, aps, pra, letterdraft, notitlepage,longbibliography]{revtex4-1} 

\usepackage{graphicx,graphics,epsfig,subfigure,times,bm,bbm,amssymb,amsmath,amsfonts,amsthm,mathrsfs,MnSymbol}
\usepackage{float}
\usepackage[matrix,frame,arrow]{xypic}
\usepackage[normalem]{ulem}
\usepackage{slashed}
\usepackage{color}
\usepackage{bm}
\usepackage{caption}
\usepackage[usenames,dvipsnames,svgnames,table]{xcolor}
\usepackage[english]{babel}
\definecolor{darkblue}{rgb}{0.0,0.0,0.3}
\usepackage[colorlinks=true,
            linkcolor=red,
            urlcolor= darkblue,
            citecolor=blue]{hyperref}

\newcommand{\bes} {\begin{subequations}}
\newcommand{\ees} {\end{subequations}}
\newcommand{\bea} {\begin{eqnarray}}
\newcommand{\eea} {\end{eqnarray}}

\def\>{\rangle}
\def\<{\langle}

\newcommand{\bra}[1]{\langle#1|}
\newcommand{\ket}[1]{|#1\rangle}

\begin{document}

\title{Optimizing Quantum Annealing Schedules with Monte Carlo Tree Search enhanced with neural networks}

\author{Yu-Qin Chen}
\affiliation{Tencent Quantum Laboratory, Tencent, Shenzhen, Guangdong, China, 518057}

\author{Yu Chen}
\affiliation{The Chinese University of Hong Kong, Hong Kong}

\author{Chee-Kong Lee}
\affiliation{Tencent Quantum Laboratory, Tencent, Shenzhen, Guangdong, China, 518057}

\author{Shengyu Zhang}
\affiliation{Tencent Quantum Laboratory, Tencent, Shenzhen, Guangdong, China, 518057}

\author{Chang-Yu Hsieh}
\email{kimhsieh@tencent.com}
\affiliation{Tencent Quantum Laboratory, Tencent, Shenzhen, Guangdong, China, 518057}

\begin{abstract}
Quantum annealing is a practical approach to approximately implement the adiabatic quantum computational
model under a real-world setting. The goal of an adiabatic algorithm is to prepare the ground state of a problem-encoded Hamiltonian at the end of an annealing path. This is typically achieved by driving the dynamical
evolution of a quantum system slowly to enforce adiabaticity. Properly optimized annealing schedules often significantly accelerate the computational process. Inspired by the recent success of deep reinforcement
learning such as DeepMind’s AlphaZero, we propose a Monte Carlo Tree Search (MCTS) algorithm and its enhanced version boosted with neural networks, which we name QuantumZero (QZero), to automate the
design of annealing schedules in a hybrid quantum-classical framework. Both the MCTS and QZero algorithms
perform remarkably well in discovering effective annealing schedules even when the annealing time is short for the 3-SAT examples we consider in this study. Furthermore, the flexibility of neural networks allows us to
apply transfer-learning techniques to boost QZero’s performance. We demonstrate in benchmark studies, that
MCTS and QZero perform more efficiently than other reinforcement learning algorithms in designing annealing schedules.
\end{abstract}

\maketitle

Quantum technologies have been advancing at an incredible pace in the past two decades. Notable achievements include the implementations of adiabatic quantum algorithms using quantum annealers. Highly non-trivial and industrially relevant applications, such as various constraint optimization problems, integer factorization  \cite{jiang2018factor}, quantum simulations  \cite{king2018observation,harris2018phase}, and quantum machine learning  \cite{willsch2020support,mott2017solving,li2018quantum}, have all been experimentally demonstrated.  Despite these initial successes, much works remain to be done to enable large-scale computations with quantum annealers. In particular, better connectivity among qubits, error and noise suppressions, engineering non-stoquastic Hamiltonians  \cite{nonstoquastic}, and optimization of annealing schedule  \cite{Herr2017OptimizingSF,zeng2016schedule} (including inhomogeneous driving  \cite{susa2018exponential} of individual qubits) are some of the pressing challenges for adiabatic quantum computations (AQC)  \cite{Albash2018,hauke2019perspectives,quant-ph/0001106,Farhi2001,article,Childs2001,quant-ph/0405098, susa2021variational, herr2017optimizing, schiffer2021adiabatic,boixo2009eigenpath}.

In this work, we address one of these challenges by proposing automated designs of annealing schedules using the Monte Carlo Tree Search (MCTS)  \cite{Coulom2007,10.1007/11871842_29,Kocsis2006,ChangShingLee2009}, and its enhanced version incorporating neural networks (NNs) to further improve the performance. This enhanced version, named QZero, is inspired by the recent success of DeepMind's AlphaZero  \cite{Silver2017,Silver2018} in mastering the game of Go. The proposed methods share many similarities with the design principles of hybrid quantum-classical algorithms for quantum circuits in the NISQ era \cite{Peruzzo2014,Kandala2017,1411.4028,McClean2016,Preskill2018,Cao2019,1905.03150}, especially a related work \cite{dalgaard2020global} that implements a deep quantum exploration version of the AlphaZero algorithm for control problems  and achieves substantial improvements in both the quality and quantity of good solution clusters compared to earlier methods. 
In fact, both approaches can be viewed more broadly as  examples of computer-automated experimental designs. A classical subroutine iteratively revises its design of annealing schedules or gate parameters, such that an annealer or a circuit may generate a desired quantum state. This classical subroutine solves an optimization problem with either a gradient-based approach (as commonly adopted in the training of neural networks) or a gradient-free optimizer such as the Bayesian approach, genetic algorithm and evolution strategy.  Recently, proposals based on reinforcement learning (RL)  \cite{zhang2019reinforcement,ChunlinChen2014,Bukov2018,Niu2019,1908.08054,1911.04574,1812.10797,hh,ayanzadeh2020reinforcement,1911.04574} to automate the experimental designs have also emerged as popular alternatives.  Conceptually, RL  \cite{10.5555/3312046,Kaelbling1996,vanOtterlo2012,1312.5602,Mnih2015,schulman} is a machine learning method that learns to accomplish tasks by interacting with an environment as opposed to simply extracting useful patterns from static data. This type of learning process makes RL to perform more robustly (in comparison to other ML methods) in a noisy and inherently stochastic environment. RL algorithms have been used in many scientific and engineering fields to address difficult problems after witnessing the remarkable accomplishments of AlphaGo and AlphaZero. Yet, we have not seen attempts in adopting MCTS, which is another indispensable ingredient for AlphaGo and AlphaZero, to automate design of annealing schedules. In fact, the underlying search mechanism of MCTS can be viewed as a learning algorithm for Markov Decision process, the central model in RL; therefore, MCTS can perform similar tasks like other RL algorithms \cite{Vodopivec2017}.  In this work, we adopt MCTS and modify the standard AlphaZero to design optimal annealing schedules.

Under the AQC paradigm, a computational problem is framed in such a way that the desired solution corresponds to the ground state of a problem-specific Hamiltonian $H_{final}$. Quantum annealing is a heuristic approach to prepare the desired ground state. 
Typically, the approach begins by initializing a quantum annealer in the ground state of a simple Hamiltonian $H_{init}$ (assuming this task can be accomplished efficiently). Next, one slowly tune the Hamiltonian towards $H_{final}$. If the dynamical process proceeds slowly enough to largely avoid Landau-Zener transitions to excited states, the adiabatic theorem should be applicable.  At the end of the annealing process, the quantum annealer should successfully prepare the ground state of $H_{final}$ with high probability. In practice, however, the annealing time cannot be arbitrarily long due to detrimental noises lurking in the background, and the fact that we expect quantum computations to be fast.  These conflicting requirements on annealing time constitute a real challenge to keep the quantum annealer in the instantaneous ground state of a time-dependent Hamiltonian with high probability. The difficulty of maintaining the adiabatic condition aggravates tremendously with the problem size; and it becomes crucial to optimize the annealing schedule \cite{Herr2017OptimizingSF,zeng2016schedule} in order to improve performance.
\onecolumngrid

\begin{figure}[htp]
\centering
\includegraphics[width=0.8\textwidth,height=0.5\textwidth]{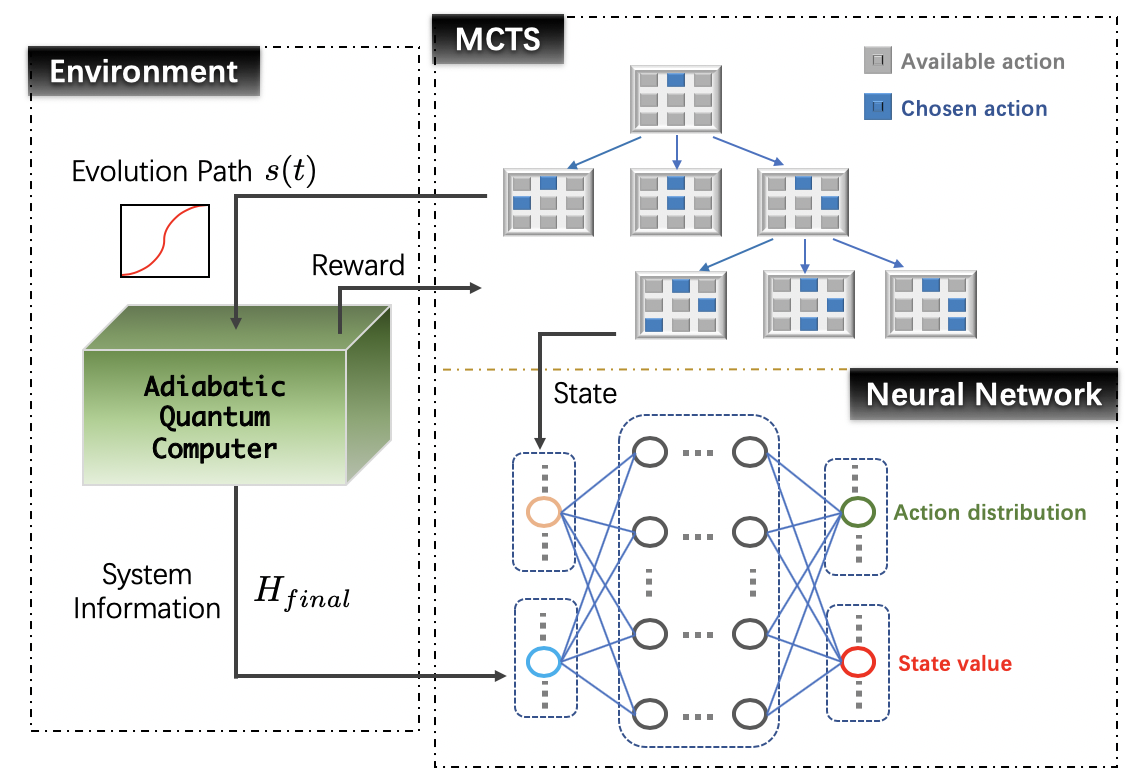}
\caption{Hybrid quantum-classical framework for designing annealing schedules. (i) Environment: a quantum annealer executes an annealing schedule encoding a specific problem and provides feedback, upon energy measurement, to a learning agent composed of MCTS and neural networks (for Quantum Zero). (ii) MCTS: the main search component of a learning agent. The search algorithm repeats the steps of selection, expansion, simulation and back propagation as introduced in the Method section. (iii) QZero:  The self-play of MCTS can be assisted with neural networks, which takes the current path explored by the MCTS as ‘state’ and ‘system information’ (the $H_{final}$) as inputs and gives out “action distribution” and “state values” as outputs to guide MCTS. These neural networks can be pre-trained as detailed in the Method section.}
\label{fig:framework}
\end{figure}

\twocolumngrid
In this study, we carefully benchmark how the MCTS performs against other RL algorithms in designing annealing schedules.  First, we elucidate the advantages shared by MCTS and other RL methods in solving difficult optimization.  As gradient-free methods, MCTS and other RL models mitigate the issues of local-minima trapping in a high-dimensional energy landscape. Secondly, both methods can efficiently handle combinatorial problems involving discrete variables. However, we hypothesize that MCTS should be a more suitable method than other RL techniques for automating quantum experiments (and quantum algorithmic designs) when it is expensive to generate high volume of training data. This hypothesis has been positively validated in our numerical study. The MCTS uses an order of magnitude less queries in finding optimal solution as manifested in the comparison of training efficiency of various algorithms in Fig.\ref{fig:eff}.  

Another major distinction between our proposed methods and prior approaches is our treatment of transfer learning.  Transfer learning skill is both remarkable and extremely useful  for acquiring optimal solutions efficiently when switching from one scenario to another, as has been studied in many other works \cite{nautrup2019optimizing, kanno2019many}.  
In the case of AlphaZero, once an RL agent is trained to devise strategy under the environment of Go, it is only expected to apply the same strategy over and over again.  However, in the context of AQC, every optimization problem is embedded in a different Hamiltonian, resulting in a different learning environment for an RL agent to learn how to prepare the corresponding ground state. While there are meta-learning strategies allowing an RL agent to adapt to different environments, it typically requires even more training time and data. In this work, we propose to simply pre-train QZero's value and policy NNs with a small set of sample problems (solved by the MCTS) such that one only needs to fine tune the NNs when the algorithm is applied to a new problem.


Finally, the proposed MCTS approaches for designing annealing schedule may be ported to the quantum circuit model \cite{1908.08054}.  By drawing the analogy between QAOA  \cite{1411.4028,1812.10797,1909.03123} and digitized quantum annealing, it is straightforward to build this connection; we leave detailed discussion in Supplementary Information. Looking more broadly, we also argue these methods can be generalized for the automated designs for other quantum technologies. Some examples include the quantum control \cite{caneva2011chopped}, quantum error corrections \cite{fosel2018reinforcement,nautrup2019optimizing}, quantum metrology \cite{xu2019generalizable}, quantum optics and quantum communications \cite{wallnofer2019machine}.

\section{Quantum Annealing and 3-SAT problem}

We first  introduce the essential background of AQC model, and elucidate how the design of an annealing schedule can be automated under the RL framework. Next, we present a constrained optimization problem, 3-SAT, used to benchmark algorithms in this work. 

\subsection{Annealing schedule as a problem for optimal control}

Quantum annealers are typically used to solve problems under the AQC framework, which relates the solutions of a problem to the ground states of a problem-encoded Hamiltonian $H_{final}$.  Preparing the ground state of an arbitrary Hamiltonian is not a simple task.  A common approach is to prepare the ground state of an alternative Hamiltonian $H_{init}$ that we can experimentally achieve with high success probability.  Next, we slowly tune the time-dependent Hamiltonian $H(s)$, along a pre-defined annealing path, towards $H_{final}$ at the end.  According to the adiabatic theorem, the time-evolved wave function will be highly overlapped with the instantaneous ground state of $H(s)$. Hence, one expects to retrieve the correct solution at the end of an annealing process with high probability.  More precisely, in each AQC calculation, we need to engineer a time-dependent Hamiltonian,
\begin{equation} \label{eq:aqc}
H(s)=(1-s)H_{init}+sH_{final},\ \  s\in [0,1].
\end{equation}
The process of tuning the Hamiltonian has to be implemented slowly in comparison to the time scale set by the minimal spectral gap of \(H(s)\) along the annealing path. Clearly, the time required to complete an AQC calculation depends crucially on the  spectral gap of $H(s)$. In reality, it is often necessary to finish the calculation within a finite duration $T$ due to various reasons such as expected quantum speedup and minimization of noise-induced errors. This time constraint (on annealing) may violate the adiabatic evolution strictly required by AQC.  Nevertheless, one can still run a quantum annealer with some schedule $s(t)$, hoping to reach the ground state of $H_{final}$ with high probability. We note this task of optimizing the schedule $s(t)$ may be framed as an optimal control problem aiming to minimize the energy as the cost function,
\begin{equation}\label{eq:aqc-min}
\begin{aligned}
\underset{{\{\mathrm{s}(\mathrm{t})\}}}{\text{argmin}}\left\langle\psi(T)\left|H_{final}\right| \psi(T)\right\rangle, 
\end{aligned}
\end{equation}
where $\{s(t):t\in [0,T]\}$ governs the state evolution  $\{|\psi(t)\rangle: t\in [0,T]\}$ through Schrodinger Equation\(\quad \frac{\partial}{\partial t}|\psi(t)\rangle=- i H(s(t))|\psi(t)\rangle\), with the starting state $|\psi(0)\rangle$, the ground state of $H_{init}$.  We remark that the adiabaticity along the annealing path is not directly reflected or assumed in the cost function, which only depends on the expected energy of the final state, $\ket{\psi(T)}$. By solving the optimal control problem above, it is likely that an optimal solution would entail a wave function $\ket{\psi(t)}$ that significantly deviates from the instantaneous ground state along a portion of the annealing path.  Usually, guided by the adiabatic theorem, it is desirable to follow the adiabatic trajectories to prepare the ground state of $H_{final}$.  Yet, it has been recently pointed out that arbitrarily long annealing time does not strictly translate into high success probability for certain problems.  Quantum annealers, operated under a finite duration $T$, may invoke diabatic transitions and yield better performances  
  \cite{karanikolas2018improved}  as observed in D-Wave experiments  \cite{king2019jp}. 

In this work, we propose a hybrid quantum-classical framework utilizing reinforcement learning (partly inspired by MCTS and AlphaZero) to design an optimal schedule $s(t)$. Fig.\ref{fig:framework} gives an overview of the proposed methods. In short, we run a quantum annealing experiment with a candidate schedule $s(t)$ and feed the result back to the MCTS-based agent in order to adjust and identify better annealing schedules in an iterative fashion.  Further details on how we adapt standard MCTS and AlphaZero for the present problem may be found in the method section.

\subsection{3-SAT problem}
In this work we use 3-SAT problems to benchmark algorithms. It is a paradigmatic example of a non-deterministic polynomial (NP) problem  \cite{Hogg2003}. A 3-SAT problem is defined by a logical statement involving \(n\) boolean variables \(b_i\). The logical statement consists of \(m\) clauses \(C_i\) in conjunction: \(C_{1} \wedge C_{2} \wedge \cdots \wedge C_{m}\). Each clause is a disjunction of 3 literals, where a literal is a boolean variable \(b_i\) or its negation \(\neg b_i\). For instance, a clause may read \(\left(b_{j} \vee \neg b_{k} \vee b_{l}\right)\). The task is to first decide whether a given 3-SAT problem is satisfiable; if so, then assign appropriate binary values to satisfy the logical statement.  

We can map a 3-SAT problem to a Hamiltonian for a set of qubits. Under this mapping, each binary variable $b_i$ is represented as a qubit state. Thus, an \(n\)-variable 3-SAT problem is mapped into a Hilbert space of dimension \(N = 2^n\). Furthermore, each clause of the logical statement is translated to a projector, that projecting on the bitstrings that not satisfying each given clause. Hence, a logical statement with $m$ clauses may be translated to the following Hamiltonian,
\begin{equation}
H_{final}=\sum_{\alpha=1}^{m}\left|b_{j}^{\alpha} b_{k}^{\alpha} b_{l}^{\alpha}\right\rangle\left\langle b_{j}^{\alpha} b_{k}^{\alpha} b_{l}^{\alpha}\right|.
\end{equation}
This Hamiltonian is diagonal in the computational basis, and the spectrum has a unit gap between eigenvalues. Each of the $m$ configurations appearing in $H_{final}$ specify the violation of a clause in the logical statement. Hence, a solution only exists if the lowest eigenvalue of $H_{final}$ is zero.  One approach to drive the $n$-qubit system to the ground state of $H_{final}$ is to use a quantum annealer under the AQC framework.

Next, we briefly mention other details essential to reproduce the numerical results in this work. Following the standard convention, we choose \(H_{init}\) for the quantum annealing algorithm to be a sum of one-qubit Hamiltonians \(H_i\) acting on the  $i$-th qubit:

\begin{equation}
H_{init}=\frac{1}{2} \sum_{i=1}^{n} h_{i} \otimes \mathbbm{1}, \quad h_{i}=\left(\begin{array}{cc}{1} & {-1} \\ {-1} & {1}\end{array}\right)
\end{equation}
The ground state of $H_{init}$ has zero energy, i.e. \(E_{0}=0\), and is a uniform superposition of all computational states which can be easily prepared by a quantum annealer.

Since the computational complexity is defined in terms of the worst-case performance, hard instances of 3-SAT have been intensively studied in the past. Following Ref.\cite{nidari2005}, we focus on a particular set of 3-SAT instances, each is characterized with a unique solution and a ratio of $m/n=3$  in this work. We note that this ratio of $3$ is different from the phase-transition point $m/n \approx 4.2$  \cite{Kirkpatrick1994,Monasson1999} that has been intensively explored in studies that characterize the degrees of satisfiability of random 3-SAT problems. The subtle distinction is that the phase-transition point characterizes the notion of ``hardness" (with respect to the $m/n$ ratio) by averaging over 3-SAT instances having variable number of solutions. However, when the focus is to identify the most difficult 3-SAT instances having unique solution, it has been ``empirically" found that these instances tend to have an $m/n$ ratio lower than the phase-transition point.

\section{Results}
In this section,  we describe several numerical experiments to illustrate the strengths of our proposed methods.

\begin{figure}[htp]
\centering
\includegraphics[width=0.44\textwidth]{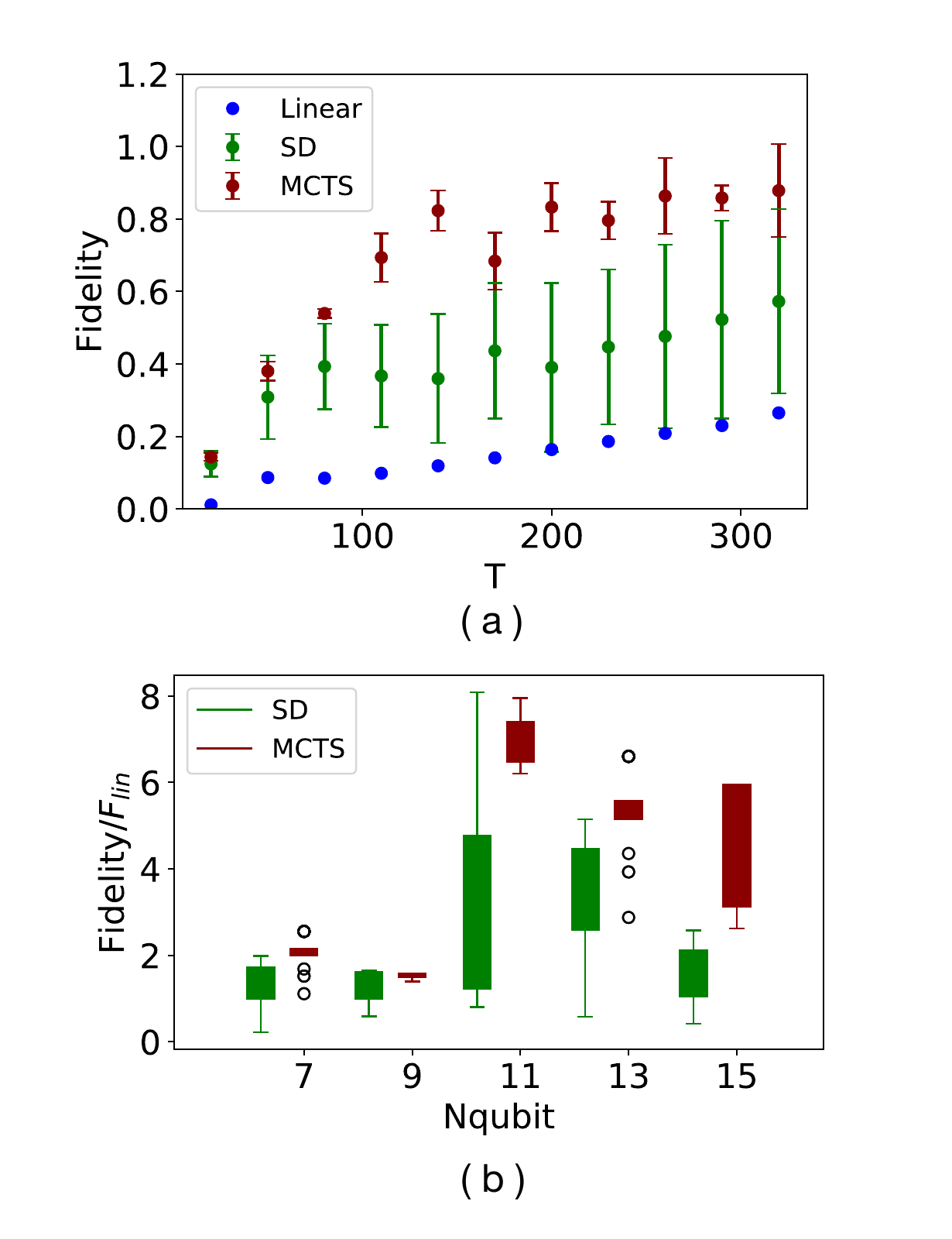}
\caption{(a)The fidelity (or success probability) of obtaining the ground states for 3-SAT instances (composed of $n=11$ variables) in a quantum annealer evolved under different annealing duration $T$.  The error bars denote the statistical fluctuations of SD and MCTS results. (b) The success probability of obtaining the ground states for 3-SAT instances (composed of $n=7,9,11,13,15$ variables) in a quantum annealer evolved under various annealing duration $T=25,40,200,300,1000$, respectively.   Here we present the results as relative values of success probability compared to  linear path $F_{lin}$.}
\label{fig:fidelity}
\end{figure}

\subsection{MCTS-designed annealing schedules}

We explain the MCTS-based automated design of annealing schedules for 3-SAT examples. MCTS is extremely efficient at solving high-dimensional optimization problems, the details of which can be seen in the method section. Usually, a proper choice of the search space may significantly simplify the process.  We elaborate on three different domains in which one may formulate the search problem as presented in the method section. In this work  we mainly focus on the design of $s(t)$ in the frequency domain as detailed in Eq.~\ref{eq:st}. 

Following Eq.~\ref{eq:st},  the goal is to pick  a sequence of \(\{x_1,x_2,x_3....x_M\}\) to minimize the energy with respect to $H_{final}$ at the end of an annealing path. 
As specified in Eq.~\ref{eq:st}, each $x_i$ corresponds to the amplitude of a frequency component when $s(t)$ is decomposed into a Fourier sine series in the $[0,T]$ domain. Since MCTS is a search algorithm, we consider $x_i$ to assume only discretized values of  \([-l_i,-l_i+\Delta_i,...,l_i-\Delta_i,l_i]\) where $l_i$ and $\Delta_i$ are some user-defined boundary value and discretization step, respectively. There is a total of \(\prod_{i=1}^M (2l_i/\Delta_i +1 )\) options of $\{x_1,x_2,x_3,\cdots x_M\}$ for the MCTS algorithms to explore. For simplicity, we set $l_i =l$ and $\Delta_i = \Delta$ in this study. In particular, we should compare path designed by our MCTS algorithm with the stochastic descent (SD) \cite{Bukov2018}, a greedy method targeting local minima in the energy landscape. The modified MCTS algorithm is presented in the Method section, while SD algorithm is briefly explained in Supplementary Information.

When the overall annealing time $T$ is sufficiently large with respect to the timescale set by the minimal spectral gap along a given annealing path, almost any schedule (including the linear one, i.e. setting $x_i=0$ in Eq:\ref{eq:st} leads to satisfactory solution with the annealer-prepared quantum state $\ket{\Psi(T)}$ having a high overlap with $\ket{\Psi_{gs}}$, the ground state of $H_{final}$.  When the annealing time $T$ is not sufficiently long, linear schedule starts to fail since Landau-Zener transitions are likely to take place when the system passes through the minimal-gap regime.  However,  resorting to methods such as MCTS or SD, it is still possible to recover non-linear schedules that significantly suppress the diabatic transitions when the tuning of the time-dependent Hamiltonian operates at a reduced rate around the critical point of minimal gap. 
We should also note the low-end regime: when $T$ is further reduced below the threshold of quantum speed limit (QST) \cite{Bukov2018}, the quantum annealer is no longer controllable, i.e. no way to attain perfect fidelity at the end of an annealing process. Since we deal with 3-SAT instances with unique solutions in this study, designing optimal annealing schedule is exactly the same as the optimal control for the state-to-state transition. As discussed in  \cite{Bukov2018}, the infidelity for state preparation (as  a function of control parameters $x_i$) transforms to a correlated phase with many non-degenerate local minima scattering around a rugged landscape. Clearly, finding global optimum (without perfect fidelity) becomes extremely difficult in this regime, $T<T_{QST}$. 
\onecolumngrid

\begin{figure}[htp]
\centering
\includegraphics[width=0.8\textwidth]{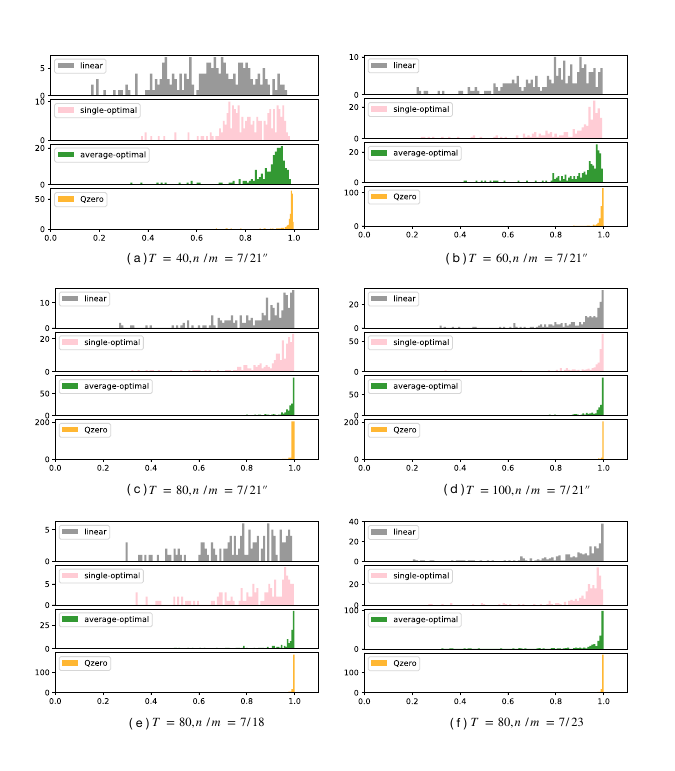}
\caption{Illustration of transferring annealing schedules across 3-SAT instances from $n/m=7/21$ to $n/m=7/21^{''}$ with annealing duration (a) \(T=40\), (b) \(T=60\), (c) \(T=80\), (d) \(T=100\). In all panels, the x-axis is the success probability and the y-axis is the number of cases.  A total of 280 examples are considered. Color codes for different results are explained in the text. llustration of transferring annealing schedules across 3-SAT instances from $n/m=7/21$ to (e) $n/m=7/18$ and (f) $n/m=7/23$ with annealing duration  \(T=80\). In all panels, the x-axis is the success probability and the y-axis is the number of cases.  A total of 280 examples are considered. Color codes for different results are explained in the text.}
\label{fig:transfer}
\end{figure}

\twocolumngrid
While the proposed annealing schedules are no longer characterized by adiabatic evolutions, the benchmarks in this section still meaningfully manifest the capability of each algorithm in solving the challenging optimization problems. Before we present our results, we describe the benchmark procedures that we consider as a fair comparison between MCTS and SD.  When solving a 3-SAT instance, MCTS has to perform many rounds of 'simulations' as it explores the control space of $x_i$  and learns to estimate the likelihood a particular annealing schedule being an optimal one. Each instance of this explorative simulation requires feedback from a quantum annealing experiment with a particular annealing path.  In comparison, every SD local search (randomly initialized with $x_i$) quickly gets stuck in a local minimum in this difficult regime.  We argue it is not fair to compare one run of MCTS search with one run of SD search, as the SD tends to query the quantum annealer significantly more times than MCTS in one run.  Rather, we will repeat SD many times (initialized with different $x_i$) such that the total number of access to a quantum annealer is comparable to that in one MCTS search.


In Fig.\ref{fig:fidelity}(a), we present the success probability of solving several 3-SAT instances of the same structure, $n=11$ and $m=33$, under different annealing durations \(T\). In this study, we fix the number of Fourier components \(M=5\),  bound strength of each Fourier component by \(l=0.2\), and set the discretization interval \(\Delta=0.01\). 
The blue points represent fidelity (or the success probability) of simple linear schedules of different annealing durations. The green points represent the average fidelity of 40 SD search with random initial conditions. The green lines give the error bars associated with SD searches. The red points represent the average fidelity of 80 episodes of a single MCTS search.  A single run of SD requires roughly 100 queries to the quantum annealers for energy feedback. On the other hand, an episode of MCTS requires roughly 50 such queries. Thus, to make a fair comparison in terms of queries to quantum annealers, we consider twice as many MCTS episodes as SD runs,  i.e.  (\(40*100=80*50\)).
According to Fig.\ref{fig:fidelity}(a), those large error bars of SD indicate a complex optimization landscape comprising multiple local minima, where SD easily gets stuck into.  On the other hand, using roughly the same number of queries to a quantum annealer, the solutions found by MCTS achieve higher successful probability.

In Fig.\ref{fig:fidelity}(b), we present the success probability of solving several 3-SAT instances with different structures, \(n=7, m=21; n=9, m=27; n=11, m=33; n=13, m=39; n=15, m=45\), under relatively short annealing times: \(T=25,T=40,T=200,T=300, T=1000\), respectively. The results here are presented as relative values of success probability under linear path. We plot successful probability of finding ground state of SD and MCTS by boxplot, which is a systematic way of displaying the data distribution based on five indicators: “minimum”, the first quartile (Q1), median, the third quartile (Q3), and “maximum”. Comparisons in Fig.\ref{fig:fidelity}(b) are again based on having almost the same number of queries to the quantum annealers as explained in the previous paragraph. As shown in the comparisons, when the optimization landscape features many local minima, local method such as SD has a high probability to get stuck, yet global method  MCTS shows the resilience and has a better chance to escape from these traps. Especially, as the problem size gets larger the optimization landscape is very likely to become more rugged, the performance gap widens between MCTS  and SD.  For instance, see \(n=11, n=13, n=15\) in Fig.\ref{fig:fidelity}(b).


\subsection{Transfer of annealing schedules}
As demonstrated in the previous section, MCTS gives higher-quality solutions than  SD, which holds even if SD is given multiple chances with different initial conditions to facilitate the exploration of the solution space.  Nevertheless, a single run of MCTS still requires repeated episodes to balance the trade-off of exploration and exploitation.  In near term, quantum resources are expensive, hence it is desirable to seek alternatives that could minimize dependence on a quantum annealer.  To this end, we resort to recent developments that combine MCTS with neural networks.

It is highly desirable if MCTS can learn from accumulated experiences of solving similar problems in the past.  In the field of deep learning, a similar goal is achieved for NNs via transfer learning.  For instance, NNs pre-trained on a large dataset can be easily adapted to predict properties of a small dataset.  Inspired by this flexibility of NNs, we further modify MCTS by incorporating NNs as done in Deep Mind's AlphaZero.  However, the off-the-shelf AlphaZero is not a suitable model for our purpose. For instance, AlphaZero only needs to learn to win the game of Go under one set of rules; but we need an algorithm that prepares ground state of multiple Hamiltonians (analogous to different rules for the game).  Another issue is that AlphaZero needs to find a winning strategy for a two-player game while there is no such competitions in our scenario. Several modifications are required before AlphaZero could use for quantum annealing, details these modifications can be found in the Method section. For clarify, we name the adapted method QuantumZero (QZero).


Here we investigate the effectiveness of transferring an annealing schedule learnt from  a set of training instances to a set of test instances under three different scenarios. The idea is that we first use MCTS to solve some sample instances similar to the actual problems we are interested in. The ``optimal" solution returned by the MCTS is then used in three different ways to guide the search for annealing schedules for new instances. First scenario is we simply solve one sample instance and apply the same annealing schedule to a set of test instances.  
Second scenario is to transfer an ``average optimal" annealing schedule to test instances. Here, the average-optimal annealing schedule is found by using MCTS to search for a schedule that gives highest 'average' success probability for multiple sample instances. Third scenario is we construct a training dataset out of ``optimal" solutions for sample instances in order to train the policy and value neural networks for QZero. When feeding QZero with new test instances, the QZero still conducts a few rounds of MCTS to fine tune the neural networks before settling on ``optimal" solutions.  As explained in the Method section, the pre-training of policy and value NNs is a relatively simple computational task because it is formulated as a standard supervised learning.  

In Fig.\ref{fig:transfer}(a)-\ref{fig:transfer}(d), we present numerical study on the the transferability  of ``optimal" annealing schedules across 3-SAT instances with different annealing duration $T=40,60,80,100$. We consider a sample set of 45 training cases and a test set of 280 examples; all problem instances share the same number of variables \(n=7\) and same number of clauses  \(m=21\).  For the first scenario,  in each annealing duration considered, we randomly select an MCTS-found schedule \(\mathbf{x}\) for a particular training example and apply this schedule to all test cases.  The results are plotted as pink-colored distributions in Fig.\ref{fig:transfer}(a)-\ref{fig:transfer}(d). For the second scenario, under different annealing durations, we take an average-optimal schedule (that gives the highest average success probability of all 45 sample instances) and apply it to test samples. These results are green ones in Fig.\ref{fig:transfer}(a)-\ref{fig:transfer}(d). Finally,  yellow results are given by QZero pre-trained with 45 training cases. We caution that the reported results given by QZero are obtained after a few rounds of fine tuning the neural networks. For comparisons, we also plot the results from the naïve linear schedule to all test cases under different annealing durations, see grey distributions in Fig.\ref{fig:transfer}(a)-\ref{fig:transfer}(d).  Going from long $T=100$ to short $T=40$ duration, it becomes progressively harder to achieve high success probability with the naive linear schedules. The pink results (a non-linear schedule adapted from a random instance) generally perform better than the linear schedule. This excellent transferability of a single annealing schedule is explained at the end of this section.  Next, green results given by the average-optimal annealing schedule manifests high percentage of obtaining a satisfying solution to any peculiarity associated with individual test cases.  Finally, the pre-trained QZero (yellow) gives the best results for all annealing durations. We remind that one needs to perform some light training to fine tune QZero's value and policy NNs for each test instance.

Next, we investigate transferability of annealing schedules (for a fixed annealing duration $T=80$) across 3-SAT instances having different $(n,m)$ parameters. In Fig.\ref{fig:transfer}(e),\ref{fig:transfer}(f), the applicability of transferring knowledge gained from optimal schedules for 45 training samples with \(n=7,m=21\) to 350 test samples with \(n=7,m=18\), see: Fig.\ref{fig:transfer}(e); between  45 training samples with \(n=7,m=21\)  and 350 test samples with \(n=7,m=23\), see:  Fig.\ref{fig:transfer}(f). Again, we consider three different strategies to use the knowledge obtained from the training set. The color codes in Fig.\ref{fig:transfer}(e),\ref{fig:transfer}(f) are identical to the ones in Fig.\ref{fig:transfer}(a)-\ref{fig:transfer}(d). It is obvious that the success probability using the optimal path transferred from a single training instance (pink) is higher than using a linear path (gray). In turn, the success probability of solving new test instances with the average-optimal schedule (green) is higher than that of the ``optimal" path of a single instance (pink). If we pre-train QZero's policy and value NNs, the results are again the best among all scenarios considered.  To address the concern (whether pre-train really accelerates the search) of having to fine tune QZero's NNs, we investigate the training  efficiency of QZero in the next subsection. 
\begin{figure}[htp]
\centering
\includegraphics[width=0.45\textwidth]{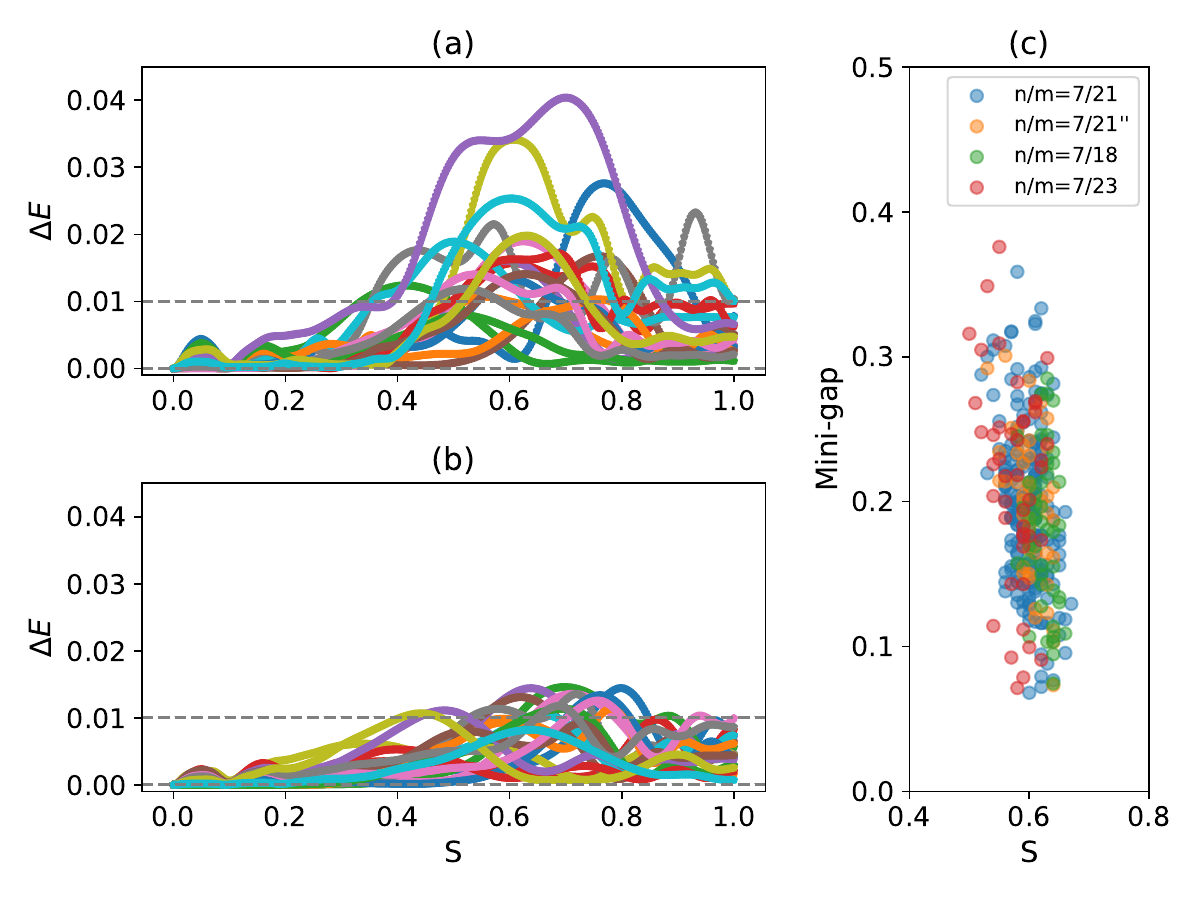}
\caption{(a) The difference between ground energy and the average energy of the time-evolved quantum state, following the SD-designed schedule, with respect to the instantaneous Hamiltonian. (b) The difference between the ground state energy of the instantaneous Hamiltonian and the average of the time-evolved quantum state, following the schedule by QZero with pre-training, with respect to the instantaneous Hamiltonian. (c) The distribution of minimal gap for 3-SAT instances used in Fig.5 ($n/m=7/21$ for training dataset, $n/m=7/21^{''}$ for test dataset) and Fig.6 ($n/m=7/18$, $n/m=7/23$).}
\label{fig:mingap}
\end{figure}

Finally, we return to the transferability of annealing schedules across 3-SAT problems. In Fig.\ref{fig:mingap}(c), the distribution of min-gaps (smallest energy gap between the first excited state and the ground state of instantaneous Hamiltonian along annealing paths) for 3-SAT instances using to produce Fig.\ref{fig:transfer}(a)-\ref{fig:transfer}(d) and Fig.\ref{fig:transfer}(e),\ref{fig:transfer}(f) is presented.  As seen, all these instances have their min-gap around $s=0.6$ with rather restricted energy range.  This high similarity of min-gap structure along different annealing paths is responsible for the high transferability of annealing schedules across instances even without sophisticated treatments as shown by the pink results in Fig.\ref{fig:transfer}(a)-\ref{fig:transfer}(d) and Fig.\ref{fig:transfer}(e),\ref{fig:transfer}(f). However, this does not imply these instances are nearly trivially identical. In Supplementary Information, we further analyze the optimal pulse profiles for some randomly chosen instances with $m/n=3$. It is clear that these pulse profiles look sufficiently distinct, which implies these instances also possess their own unique gap profiles along the annealing paths. This also explains why the transferability drops when $T=40$ is small except for the Qzero model which performs the standard transfer learning with additional learning steps to fine tune the designed path for each individual problem. 

The differences between ground energy and the expected energy of the time-evolved quantum state following SD or QZero annealing schedules are carefully investigated in Fig.\ref{fig:mingap}(a) and Fig.\ref{fig:mingap}(b), respectively. The energy difference \(\Delta E\) reflects how strongly the adiabaticity is violated along different paths. As shown, the pre-trained QZero is not only able to find optimal solutions but also to enforce adiabaticity better than SD.  


\subsection{Comparing learning efficiency of Qzero and other RL methods}
\begin{figure}[htp]
\centering
\includegraphics[width=0.45\textwidth]{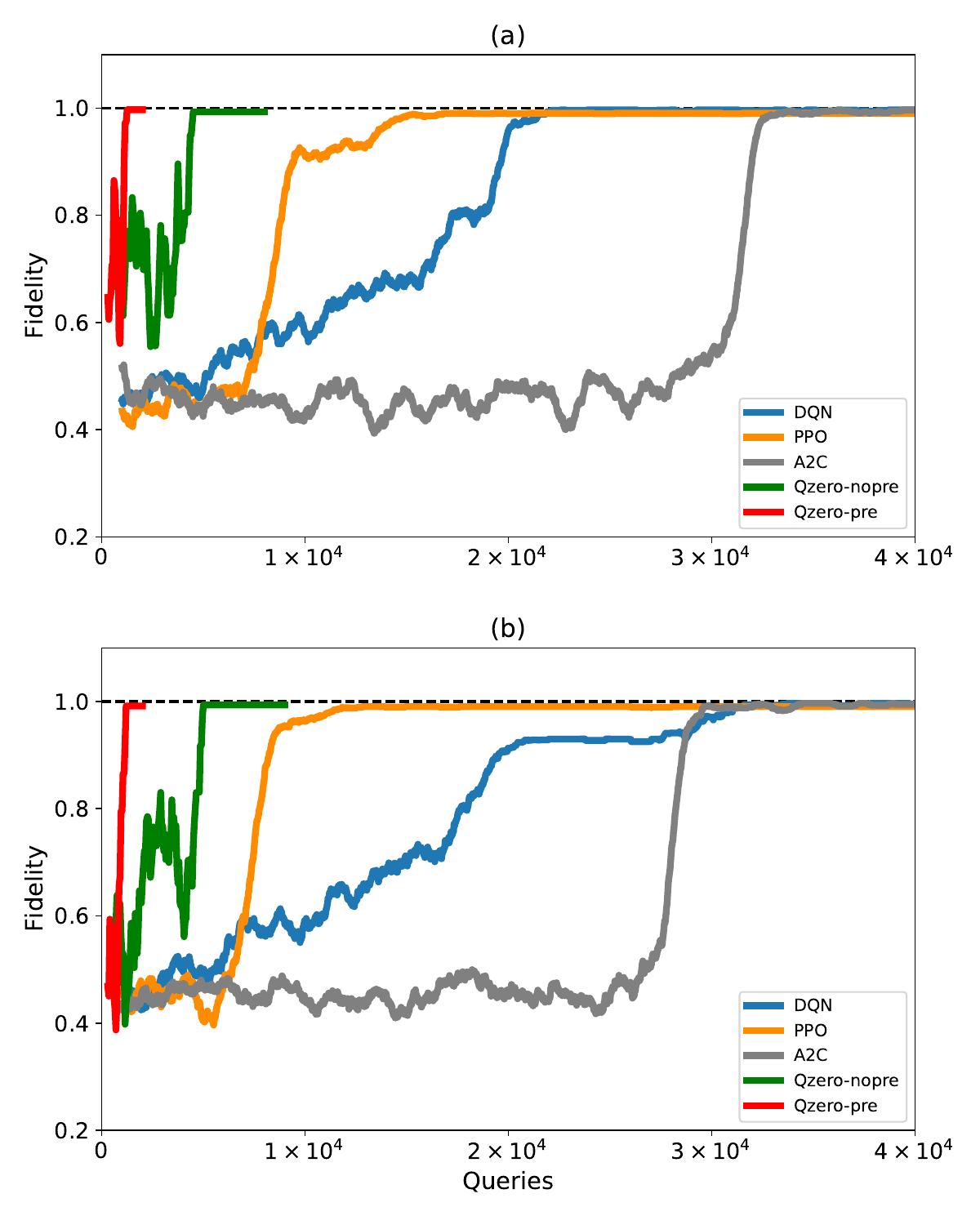}
\caption{Comparing the learning efficiency among RL algorithms. (a) Using an example Hamiltonian \(H_{final}^1\) to compare Qzero without pre-train (Qzero-nopre), Qzero with pre-train (Qero-pre) and three other RL methods: DQN, PPO, A2C.  (b) Using \(H_{final}^2\) to compare  Qzero without pre-train (Qzero-nopre), Qzero with pre-train (Qero-pre) and three other RL methods: DQN, PPO, A2C. }
\label{fig:eff}
\end{figure}

Finally, we compare the learning efficiency of Qzero with other popular RL methods mentioned in the introduction.  Similar to Qzero,  these RL methods are capable of finding global optimum even for difficult problems like the ones discussed in the previous sections. However,  training typical RL methods are notoriously resource consuming.  Here, we demonstrate that Qzero achieves the same level of performance (as other RL methods) using less computational resource. In particular, our assessment is based on  the number of queries to a quantum annealer required by each method. In this benchmark, we compare two variants of MCTS algorithms: QZero with pre-training ('QZero-pre'), QZero without pre-training ('QZero-nopre') ; and three RL models: deep Q-networks (DQN)  \cite{1312.5602, Mnih2015}, Advantage-Actor-Critic (A2C)  \cite{10.5555/3312046} and proximal policy optimization (PPO) \cite{schulman}. See Supplementary Information for details of these three RL algorithms.

We use two 3-SAT examples, denoted by \(H_{final}^1\) and \( H_{final}^2\) of  size \(n=7,m=21\), as benchmark for this efficiency test to design an annealing schedule with duration \(T=70\).  We formulate all RL algorithms to possess an identical set of actions for designing the Fourier components of all allowable schedules defined in Eq.\ref{eq:st}. In particular, we consider five frequency components,
 \(M=5\), and each coefficient \(x_i\) belongs to a discretized space of \([-l,-l+\Delta,...,l-\Delta,l]\),  where \(l=0.2\) and \( \Delta=0.01\).
The efficiency test is summarized in Fig.\ref{fig:eff}. We look at how fast each algorithm finishes its training and returns an optimal solution.
In this figure, a ``query" specifically refers to operating a quantum annealer with an annealing schedule in order to provide feedback.  To make fair comparisons, the queries 'hidden inside the simulation playouts' of QZero are explicitly taken into account. As manifested in the figure, QZero (without pre-train) performs more efficiently than all other RL methods (DQN, PPO, A2C) as the MCTS performs efficient searches. Qzero equipped with pre-trained networks gets a further boost in the learning efficiency. For additional details, please see the analysis on the convergence efficiency with respect to the system size in Supplementary Information. In addition to comparing with other RL algorithms, we also compare the performances between QZero and MCTS (the core search component in QZero). These details can be found in Supplementary Information too.


\section{Discussion}
In this work, we propose data-driven approaches to design annealing schedules for solving combinatorial problems in a quantum annealer.  These approaches build on the venerable search algorithm Monte Carlo Tree Search (MCTS) and a generalization, termed Quantum Zero (QZero), incorporating neural networks.  Since the trainings of neural networks (NNs) may take significant amount of time and computational resources, we propose to pre-train them with a collection of sample problems using a MCTS solver. These pre-trained NNs learn to transfer annealing schedules between similar problem instances.  This pre-training strategy generalizes the standard AlphaZero algorithm from interacting with one environment (corresponding to one problem instance) to efficiently adapt and interact with multiple environments.

In this study, we have demonstrated that MCTS outperforms the stochastic descent, a local search algorithm, when addressing tough problems characterized by complex energy landscape. In addition, we also compare MCTS and QZero to a host of other RL algorithms, that recently attracted significant attention because of their potential to improve quantum annealing as well as QAOA algorithms for solving the combinatorial problems.  We have found the MCTS and QZero outperform all other RL algorithms considered in our benchmark study.  In particular, the pre-trained QZero turns out to be the most efficient among all RL algorithms reported in this work. Our work shows that MCTS and Qzero are highly competitive methods for automating designs of quantum annealing schedules.

\section{Methods}

 In this section, we introduce our proposed strategy to design $s(t)$ with the Monte Carlo Tree Search and a neural-network enhanced version QuantumZero.
\begin{figure}[htp]
\centering
\includegraphics[width=0.44\textwidth]{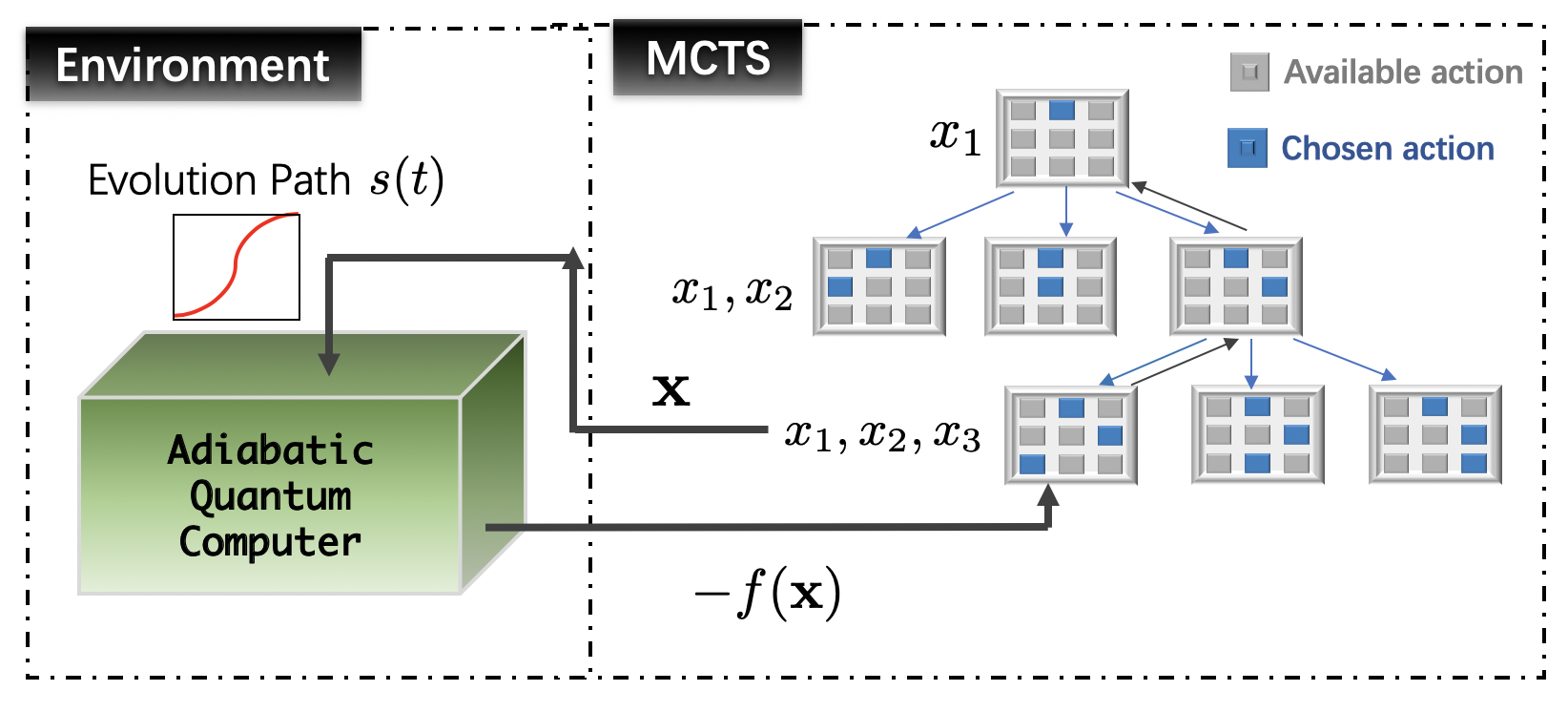}
\caption{Setup of MCTS.}
\label{fig:mcts}
\end{figure}

\subsection{Monte Carlo tree search}
 MCTS aims at finding a vector of discrete variables \(\textbf{x}^*\) that maximizes or minimizes a target property \(f(\textbf{x})\) evaluated by a 
problem-specific learning environment. For designing an annealing schedule, \(\textbf{x}=\{x_1,x_2,x_3,...,x_M\}\) corresponds to coefficients of 
Fourier series introduced in Eq.\ref{eq:st}.  Each \(x_i\in\{-l_i,-l_i+\Delta_i,\cdots,l_i-\Delta_i,l_i\}\),  \(\pm l_i\) are the upper and lower bound for 
the amplitudes of $i$-th frequency component, and \(\Delta_i\) is the discretized increment in the frequency space.  The whole search space is 
composed of \(\prod_{i=1}^M(2l_i/\Delta_i +1)\) grid points. In our case, \(f(\textbf{x})=\left\langle\psi_{\textbf{x}}(T)\left|H_{final}\right| 
\psi_{\textbf{x}}(T)\right\rangle\) is the expected energy, where $\ket{\psi_{\mathbf{x}}(T)}$ is the time-evolved quantum state at the end of an 
annealing path.

MCTS performs the search on an ($M+1$)-level tree structure. The zero-th level is just a root node, which denotes a starting point and carries no other significance. The nodes at the \(k\)-th level correspond to the $(2l_k/\Delta_k +1)$ value assignments of \(x_k\) with $k=1,\cdots,M$.  
Every solution $\mathbf{x}$ specifies a path along the tree structure from top to bottom.  In Fig. \ref{fig:mcts}, we illustrate how a MCTS search is conducted on a $(3+1)$-level tree composed of 3 nodes in each level.  Ignoring the zero-th level, the tree structure actually looks like a 3 by 3 square board shown in the right panel of Fig.~\ref{fig:mcts}. The MCTS starts at the root and traverses the tree level by level.
The algorithm has to select a node (blue box in figure) before proceeding to the next level. As illustrated in the figure, the MCTS decides a path by sequentially inserting $x_1$, $x_2$ and $x_3$  into an array specifying the path, \(\{\},\{x_1\},\{x_1,x_2\},\{x_1,x_2,x_3\}\)

Each round of MCTS consists of four stages: selection, expansion, simulation and back propagation. In the selection stage, a path is traversed from  the root down to a node $x_k$ at \(k\)-th level by choosing the nodes $x_i$ (with $i\leq k$) having maximum Upper Confidence Bound (UCB) score at each level \(x_i=\max_{a} u_a\), where the maximum is over candidate actions $a$. The UCB score indicates how promising it is to explore the subtree under the current node and is defined as
\begin{equation}
u_{a}=\frac{w_{a}}{v_{a}}+C \sqrt{\frac{2 \ln v_{\text {parent }}}{v_{a}}}
\end{equation}
where the visit count \(v_a\) denotes the number of visits to node $a$ during the search process, \(v_{parent}\) is the visit count of the parent 
node, the cumulative merit \(w_a\) is defined as the sum of all direct merits for all descendant nodes including itself, and \(C\) is a constant to balance the exploration and exploitation. This traversal terminates at $k$-th level when all its children nodes have not been visited before.  At this point, the 
search enters the expansion stage. \(N_{exp}\) new children nodes  are added under the current node \(x_k\) with following initializations: \(v_{a}=w_{a}=f_{a}=0, 
u_{a}=\infty\) relevant for the UCB score. Once new children nodes are created, the search transits to the simulation stage. \(N_{sim}\) times of  random playout are 
performed for each of the added children node. A playout is a random selection of additional nodes to form a complete path from top to bottom, the 
$M$-th level.  Once such a  path has been randomly picked,  \(f(\textbf{x})=\left\langle\psi_{\textbf{x}}(T)\left|H_{final}\right| \psi_{\textbf{x}}(T)
\right\rangle\) is evaluated and recorded as an immediate merit of the path.  In the final stage of back propagation, the visit count of each ancestor 
nodes of \(x_i\) is incremented by one and the cumulative value is also updated to maintain consistency.   We repeatedly run this 4-stage search for a 
fixed number of times. At the end, the best solution would be returned as the final result. The random playouts of MCTS allow us to efficiently explore a large 
set of candidate solutions, and identify promising directions to focus the search for optimal solutions. 


For experiments reported in the main text, we set the constant  \(C=2\) to balance the exploration and exploitation, the number of nodes added at each expansion \(N_{exp}=10\), and the simulation times at a node \(N_{sim}=5\).  

\subsection{Search Space for Annealing Schedules}

MCTS and related modern methods are extremely efficient at solving high-dimensional optimization problems that we encounter in this work. However, depending on the context of the problem at hand, a proper choice of the search space may significantly simplify the process.  In this section. we elaborate on three different domains in which one may formulate the search problem:   
                          
\begin{itemize}
  \item [1)] 
  \textbf{Time domain.} Directly designing $s(t)$ in the time domain is a straightforward idea. The optimal control problem is now turned into assigning values (from a predefined range) to a sequence \(\mathbf{x}=\{x(0),x(\delta t),x(2\delta t)....x(T)\}\) after uniformly dividing the evolution time $T$ into small segments of $dt$.  
  \item [2)]
  \textbf{Frequency domain.}  \(s(t)\) can also be Fourier expanded around a monotonically increasing schedule, such as $s_0(t)=t/T$, in the frequency domain:
\begin{equation}\label{eq:st}
\mathrm{s}(\mathrm{t})=s_0(t) +\sum_{i=1}^{M} x_{i} \sin \frac{i \pi t}{T},
\end{equation}
with $M$ the total number of Fourier components. The optimal control problem is now reduced to assigning values to the sequence \(\mathbf{x}=\{x_1,x_2,x_3....x_M\}\). This Fourier series expansion is a common approach to expand the pulse profile in some truncated functional basis set. For instance, a mainstream quantum optimal control scheme, Chopped Random Basis (CRAB ) ~\cite{caneva2011chopped} method, also employs similar expansion but with randomized frequency components. We expect our method can be easily integrated with CRAB and other optimal control methods to serve more potential applications in the domain of quantum technology.

\item [3)]
  \textbf{Hybrid Time-Frequency Domain.}  This is a generalization combining the strengths of pulse designs in both time and frequency domains.  One such approach is to optimize not only Fourier coefficients in Eq.~(\ref{eq:st}) but also $s_0(t)$.  For instance, a promising method is to incorporate the bang-bang control into $s_0(t)$, as a recent study suggests that an optimal control schedule should assume a bang-anneal-bang profile \cite{brady2021optimal} for the quantum circuit model solving the same kind of combinatorial problems discussed in this work. It is obvious that the numerical optimization becomes extremely challenging if we treat the high-frequency (due to the fast bang-bang flipings) part of $s_0(t)$ and the Fourier sine expansion part in a unified Fourier analysis in the frequency domain. It will be more convenient if we refine the pulse-design strategy accordingly.  See Supplementary sec.D for more details on a test study wherein we impose $s_0(t)$ to take on a linear schedule for the most part but switch to a bang-bang control for a brief interval at the beginning and the end of the schedule. Additionally, one may also expand and optimize the highly non-smooth schedule $s(t)$ in a wavelet basis due to its superior capacity to model multi-scale design problems.  
\end{itemize}

In this study, we mainly focus on the design of $s(t)$ in the frequency domain as detailed in Eq.~(\ref{eq:st}). As manifested in our simulation experiments on 3-SAT problems, MCTS-guided search in the frequency domain already exhibits superior performances to other conventional methods. Therefore, in the main text, we do not further investigate the effectiveness of  conducting MCTS-guided search in the hybrid time-frequency domain. 

\subsection{QuantumZero}
While MCTS is an extremely powerful approach to search a large combinatorial space, it is nevertheless a time-consuming procedure especially, when the space grows exponentially with the number of Fourier components. If one is expected to solve a large set of similar problems, it will be highly desirable that one can utilize past experiences in solving similar problems to accelerate the search.  One way to achieve this goal is to combine MCTS with NNs and resort to a host of transfer-learning techniques.  Inspired by the design of AlphaZero, we introduce both policy and value NNs to enhance search efficiency of MCTS. Furthermore, these NNs can be straightforwardly pre-trained by learning from past experiences.  Below, we discuss how we modify the standard AlphaZero for quantum annealing. The following three points highlight the main differences between DeepMind's AlphaZero and our modified algorithm QZero.

(1) QZero is a single-player game without competition. The win (or loss) of a QZero game is determined by the satisfaction (or dissatisfaction) of this inequality,  \(E-E_g<\epsilon\) where \(E=\left\langle\psi_{\textbf{x}}(T)\left|H_{final}\right| \psi_{\textbf{x}}(T)\right\rangle\) and $E_g$ is the ground-state energy of $H_{final}$.

(2) AlphaZero only deals with a single chessboard as the learning environment.  In order to facilitate transfer learning between different  environments across problem instances, QZero's NNs require input information regarding a specific Hamiltonian \(H_{info}\). Take a 3-SAT instance with \(n\) variables and \(m\) clauses for example, \(H_{info}\) is an \(m\times n\) matrix encoding information about all clauses. Variables \(b_i\) is encode as 1,  its negation \(\neg b_i\) is encode as -1. For \(s-\)th clause \(\left(b_{j} \vee \neg b_{k} \vee b_{l}\right)\), we have \(H^{s,j}_{info}=1, H^{s,k}_{info}=-1, H^{s,l}_{info}=1\) and \( H^{s,others}_{info}=0\). Then we deform \(H_{info}\) into vectorized form, \(\vec{H}_{info}\), which is then attached to the vector of chessboard state as input to the NNs. 

(3) The NNs can be efficiently pre-trained with datasets processed by MCTS. 
The pre-train dataset has the structure, $\{\vec{s}^i,\vec{p}^i,\vec{v}^i \vert i=1,\dots,N_{sample}\}$. For the $i$-th instance, the input data reads $\vec{s}^i =  \{(0,0,0,...), (x_1,0,0,0..),..(x^i_1,x^i_2,...,x^i_{M-1},0) , \vec{H}^i_{info} \}$, where $x^i_k$ are components of a solution $\mathbf{x}^i$ (found by a pure MCTS search) for a sample instance given by $\vec{H}^i_{info}$. The corresponding output label $\vec{p}^i$  require more work to construct.  First, we take a vector of size $M$ from the set  \(\{ (x_1^i,0,0,0..),(0,x_2^i,0,0..),..(0,0,...,x_M^i) \}\) and convert it to a new vector of size \(M*(2l/\Delta+1)\). Again, we assume that $l_j = l$ and $\Delta_j = \Delta$ for simplicity.
Instead of directly specifying the coefficient $x_m^i$, we may just indicate which of the $2l/\Delta + 1$ choices $x_m^i$ corresponds to. 
For instance,  we create a new vector $\tilde p^i_j$ out of  \((0,x_j^i,0,0..) \) as follows,
\begin{equation}
\left\{\begin{array}{l}  \tilde p^i_{jk}=1,\ \  k=(x_j+l)/\Delta +(2l/\Delta+1)(j-1), \\ 
\tilde{p}^i_{jk}=0, \ \ \text{otherwise} \end{array}\right.
\end{equation}
The $i$-the sample data $\vec{p}^i=[\tilde{p}^i_1, \cdots, \tilde{p}^i_M]^T$ is vector assembled from concatenation of all $\tilde{p}^i_j$. 
The other output label $\vec{v}^i=\{1,1,...1\}$ carries only one value as shown.  This is because we only consider the ``winning" strategy for each sample instance in the pre-train dataset. The value and policy neural network are then trained as a supervised-learning task, 
\begin{equation}
(\vec{p}, \vec{v})=G_{\theta}(\vec{s}),
\end{equation}
where \(\theta\) is the weights of the neural network \(G\).  These pre-trained NNs can be easily incorporated into MCTS as discussed below.  

Even though QZero is pre-trained, the NNs still require fine tuning when applied to a new problem instance.
The training process  proceeds in two stages.  First, MCTS equipped with pre-trained NNs goes through a modified search procedure (same as AlphaZero) multiple times, picks new annealing schedule $\mathbf{x}_i$ each time, and obtains corresponding evaluation $v_i$ given by the learning environment. In the second stage, this set of collected data $\{\mathbf{x}_i, v_i\}$ is used to further train neural networks by following the AlphaZero algorithm. The detail is provided at the end of the next paragraph.

The 4-stage MCTS is modified to make use of the action distribution and state value estimated by NNs as a guidance for selecting path traversal. The streamlined QZero comprises of a three-step procedure: selection, expansion with evaluation, and back propagation. The selection step relies on a score function to decide a path traversal along the tree structure. The modified score function reads
\begin{equation}\label{eq:ucb2}
U_{\vec{s}, a}=\frac{W_{\vec{s}, a}}{N_{\vec{s}, a}}+C\  \vec{p}_{\vec{s}, a} \frac{\sqrt{\sum_{a'} N_{\vec{s}, a'}}}{1+N_{\vec{s}, a}},
\end{equation}
where $a$ and $a^\prime$ represent the candidate nodes (that could be appended to extend the current path $\vec{s}$) at this selection step, the visit count \(N\) represents the visit times in the search process, \(\sum{N}\) is the visit count of the parent node, the cumulative merit \(W\) is defined as the sum of all cumulative merits for its descendant nodes including itself. The direct merit here is the value \(v\) estimated by the value NN for a partial game or otherwise \(\pm1\) for a  win or loss  for a complete game. \(\vec{p}\) is the policy value given by the policy NN. C is a constant to balance the exploration and exploitation. Repeating the selection step until arriving at a leaf node, the algorithm then expands the tree to the next level. Each leaf node at the new level is evaluated by the direct merit \(v\) defined earlier, and this merit \(v\) is back propagated to update the cumulative merits \(W\) for all its parent nodes along the search tree. After \(N_{playout}\) simulations, QZero makes an actual move based on a new policy distribution $\pi$, which is updated with the frequency counts of attempted actions during simulations. An episode is finished when QZero makes a sequence of actual moves to fully specify an annealing schedule, i.e. reaching the bottom of the search tree. The feedback (whether the time-evolved state has a small enough energy at the end, a win-or-loss situation) by the quantum annealer essentially produces updated values  \(z\) for all explored partial or full annealing schedules in this episode. After playing through a fixed number of episodes, the collected set of data is then subsequently used to re-train the neural networks by minimizing the following loss function.
\begin{equation}
 l=(z-v)^{2}-\vec{\pi}^{\mathrm{T}} \log \vec{p}+\lambda\|\theta\|^{2},
 \label{eq:losss}
\end{equation}
where \(\lambda\) corresponds to the regularization strength of NN weights.
After calibrating the NNs with updated data, we carry out another round of MCTS guided by the new policy and value. By repeating this process of MCTS and calibrating NNs, a steady state could be reached, where the MCTS captures an optimal search strategy and  training of neural networks converges with the loss of Eq:\ref{eq:losss} tending to zero.


Finally, we report hyper-parameters for the QZero used in the section Result.  We initialize the constant to balance the exploration and exploitation at \(C=3\) and gradually decrease it to \(C=0.5\), the number of simulations before each move is \(N_{playout}=6\), policy NN has three dense layers of dimension $\{256, 128, 2l/\Delta*M\}$, and value NN has four dense layers of dimension $\{256, 128, 64,1\}$, learning rates for NN start at \(lr=0.008\) and gradually decay to \(l_r=0.0008\), and the energy error is set to \(\epsilon=0.01\).


\begin{thebibliography}{60}%
\makeatletter
\providecommand \@ifxundefined [1]{%
 \@ifx{#1\undefined}
}%
\providecommand \@ifnum [1]{%
 \ifnum #1\expandafter \@firstoftwo
 \else \expandafter \@secondoftwo
 \fi
}%
\providecommand \@ifx [1]{%
 \ifx #1\expandafter \@firstoftwo
 \else \expandafter \@secondoftwo
 \fi
}%
\providecommand \natexlab [1]{#1}%
\providecommand \enquote  [1]{``#1''}%
\providecommand \bibnamefont  [1]{#1}%
\providecommand \bibfnamefont [1]{#1}%
\providecommand  \citenamefont [1]{#1}%
\providecommand \href@noop [0]{\@secondoftwo}%
\providecommand \href [0]{\begingroup \@sanitize@url \@href}%
\providecommand \@href[1]{\@@startlink{#1}\@@href}%
\providecommand \@@href[1]{\endgroup#1\@@endlink}%
\providecommand \@sanitize@url [0]{\catcode `\\12\catcode `\$12\catcode
  `\&12\catcode `\#12\catcode `\^12\catcode `\_12\catcode `\%12\relax}%
\providecommand \@@startlink[1]{}%
\providecommand \@@endlink[0]{}%
\providecommand \url  [0]{\begingroup\@sanitize@url \@url }%
\providecommand \@url [1]{\endgroup\@href {#1}{\urlprefix }}%
\providecommand \urlprefix  [0]{URL }%
\providecommand \Eprint [0]{\href }%
\providecommand \doibase [0]{http://dx.doi.org/}%
\providecommand \selectlanguage [0]{\@gobble}%
\providecommand \bibinfo  [0]{\@secondoftwo}%
\providecommand \bibfield  [0]{\@secondoftwo}%
\providecommand \translation [1]{[#1]}%
\providecommand \BibitemOpen [0]{}%
\providecommand \bibitemStop [0]{}%
\providecommand \bibitemNoStop [0]{.\EOS\space}%
\providecommand \EOS [0]{\spacefactor3000\relax}%
\providecommand \BibitemShut  [1]{\csname bibitem#1\endcsname}%
\let\auto@bib@innerbib\@empty
\bibitem [{ \citenamefont {Jiang}\ \emph {et~al.}(2018) \citenamefont {Jiang},
   \citenamefont {Britt},  \citenamefont {McCaskey},  \citenamefont {Humble},\
  and\  \citenamefont {Kais}}]{jiang2018factor}%
  \BibitemOpen
  \bibfield  {author} {\bibinfo {author} {\bibfnamefont {Shuxian}\ \bibnamefont
  {Jiang}}, \bibinfo {author} {\bibfnamefont {Keith~A}\ \bibnamefont {Britt}},
  \bibinfo {author} {\bibfnamefont {Alexander~J}\ \bibnamefont {McCaskey}},
  \bibinfo {author} {\bibfnamefont {Travis~S}\ \bibnamefont {Humble}}, \ and\
  \bibinfo {author} {\bibfnamefont {Sabre}\ \bibnamefont {Kais}},\ }\bibfield
  {title} {\enquote {\bibinfo {title} {Quantum annealing for prime
  factorization},}\ }\href@noop {} {\bibfield  {journal} {\bibinfo  {journal}
  {Sci. Rep.}\ }\textbf {\bibinfo {volume} {8}},\ \bibinfo {pages} {1--9}
  (\bibinfo {year} {2018})}\BibitemShut {NoStop}%
\bibitem [{ \citenamefont {King}\ \emph {et~al.}(2018) \citenamefont {King},
   \citenamefont {Carrasquilla},  \citenamefont {Raymond},  \citenamefont
  {Ozfidan},  \citenamefont {Andriyash},  \citenamefont {Berkley},  \citenamefont
  {Reis},  \citenamefont {Lanting},  \citenamefont {Harris},  \citenamefont
  {Altomare} \emph {et~al.}}]{king2018observation}%
  \BibitemOpen
  \bibfield  {author} {\bibinfo {author} {\bibfnamefont {Andrew~D}\
  \bibnamefont {King}}, \bibinfo {author} {\bibfnamefont {Juan}\ \bibnamefont
  {Carrasquilla}}, \bibinfo {author} {\bibfnamefont {Jack}\ \bibnamefont
  {Raymond}}, \bibinfo {author} {\bibfnamefont {Isil}\ \bibnamefont {Ozfidan}},
  \bibinfo {author} {\bibfnamefont {Evgeny}\ \bibnamefont {Andriyash}},
  \bibinfo {author} {\bibfnamefont {Andrew}\ \bibnamefont {Berkley}}, \bibinfo
  {author} {\bibfnamefont {Mauricio}\ \bibnamefont {Reis}}, \bibinfo {author}
  {\bibfnamefont {Trevor}\ \bibnamefont {Lanting}}, \bibinfo {author}
  {\bibfnamefont {Richard}\ \bibnamefont {Harris}}, \bibinfo {author}
  {\bibfnamefont {Fabio}\ \bibnamefont {Altomare}},  \emph {et~al.},\
  }\bibfield  {title} {\enquote {\bibinfo {title} {Observation of topological
  phenomena in a programmable lattice of 1,800 qubits},}\ }\href@noop {}
  {\bibfield  {journal} {\bibinfo  {journal} {Nature}\ }\textbf {\bibinfo
  {volume} {560}},\ \bibinfo {pages} {456--460} (\bibinfo {year}
  {2018})}\BibitemShut {NoStop}%
\bibitem [{ \citenamefont {Harris}\ \emph {et~al.}(2018) \citenamefont {Harris},
   \citenamefont {Sato},  \citenamefont {Berkley},  \citenamefont {Reis},
   \citenamefont {Altomare},  \citenamefont {Amin},  \citenamefont {Boothby},
   \citenamefont {Bunyk},  \citenamefont {Deng},  \citenamefont {Enderud} \emph
  {et~al.}}]{harris2018phase}%
  \BibitemOpen
  \bibfield  {author} {\bibinfo {author} {\bibfnamefont {R}~\bibnamefont
  {Harris}}, \bibinfo {author} {\bibfnamefont {Y}~\bibnamefont {Sato}},
  \bibinfo {author} {\bibfnamefont {AJ}~\bibnamefont {Berkley}}, \bibinfo
  {author} {\bibfnamefont {M}~\bibnamefont {Reis}}, \bibinfo {author}
  {\bibfnamefont {F}~\bibnamefont {Altomare}}, \bibinfo {author} {\bibfnamefont
  {MH}~\bibnamefont {Amin}}, \bibinfo {author} {\bibfnamefont {K}~\bibnamefont
  {Boothby}}, \bibinfo {author} {\bibfnamefont {P}~\bibnamefont {Bunyk}},
  \bibinfo {author} {\bibfnamefont {C}~\bibnamefont {Deng}}, \bibinfo {author}
  {\bibfnamefont {C}~\bibnamefont {Enderud}},  \emph {et~al.},\ }\bibfield
  {title} {\enquote {\bibinfo {title} {Phase transitions in a programmable
  quantum spin glass simulator},}\ }\href@noop {} {\bibfield  {journal}
  {\bibinfo  {journal} {Science}\ }\textbf {\bibinfo {volume} {361}},\ \bibinfo
  {pages} {162--165} (\bibinfo {year} {2018})}\BibitemShut {NoStop}%
\bibitem [{ \citenamefont {Willsch}\ \emph {et~al.}(2020) \citenamefont
  {Willsch},  \citenamefont {Willsch},  \citenamefont {De~Raedt},\ and\
   \citenamefont {Michielsen}}]{willsch2020support}%
  \BibitemOpen
  \bibfield  {author} {\bibinfo {author} {\bibfnamefont {Dennis}\ \bibnamefont
  {Willsch}}, \bibinfo {author} {\bibfnamefont {Madita}\ \bibnamefont
  {Willsch}}, \bibinfo {author} {\bibfnamefont {Hans}\ \bibnamefont
  {De~Raedt}}, \ and\ \bibinfo {author} {\bibfnamefont {Kristel}\ \bibnamefont
  {Michielsen}},\ }\bibfield  {title} {\enquote {\bibinfo {title} {Support
  vector machines on the d-wave quantum annealer},}\ }\href@noop {} {\bibfield
  {journal} {\bibinfo  {journal} {Comput. Phys Commun.}\ }\textbf {\bibinfo
  {volume} {248}},\ \bibinfo {pages} {107006} (\bibinfo {year}
  {2020})}\BibitemShut {NoStop}%
\bibitem [{ \citenamefont {Mott}\ \emph {et~al.}(2017) \citenamefont {Mott},
   \citenamefont {Job},  \citenamefont {Vlimant},  \citenamefont {Lidar},\ and\
   \citenamefont {Spiropulu}}]{mott2017solving}%
  \BibitemOpen
  \bibfield  {author} {\bibinfo {author} {\bibfnamefont {Alex}\ \bibnamefont
  {Mott}}, \bibinfo {author} {\bibfnamefont {Joshua}\ \bibnamefont {Job}},
  \bibinfo {author} {\bibfnamefont {Jean-Roch}\ \bibnamefont {Vlimant}},
  \bibinfo {author} {\bibfnamefont {Daniel}\ \bibnamefont {Lidar}}, \ and\
  \bibinfo {author} {\bibfnamefont {Maria}\ \bibnamefont {Spiropulu}},\
  }\bibfield  {title} {\enquote {\bibinfo {title} {Solving a higgs optimization
  problem with quantum annealing for machine learning},}\ }\href@noop {}
  {\bibfield  {journal} {\bibinfo  {journal} {Nature}\ }\textbf {\bibinfo
  {volume} {550}},\ \bibinfo {pages} {375--379} (\bibinfo {year}
  {2017})}\BibitemShut {NoStop}%
\bibitem [{ \citenamefont {Li}\ \emph {et~al.}(2018) \citenamefont {Li},
   \citenamefont {Di~Felice},  \citenamefont {Rohs},\ and\  \citenamefont
  {Lidar}}]{li2018quantum}%
  \BibitemOpen
  \bibfield  {author} {\bibinfo {author} {\bibfnamefont {Richard~Y}\
  \bibnamefont {Li}}, \bibinfo {author} {\bibfnamefont {Rosa}\ \bibnamefont
  {Di~Felice}}, \bibinfo {author} {\bibfnamefont {Remo}\ \bibnamefont {Rohs}},
  \ and\ \bibinfo {author} {\bibfnamefont {Daniel~A}\ \bibnamefont {Lidar}},\
  }\bibfield  {title} {\enquote {\bibinfo {title} {Quantum annealing versus
  classical machine learning applied to a simplified computational biology
  problem},}\ }\href@noop {} {\bibfield  {journal} {\bibinfo  {journal} {npj
  Quantum Inf.}\ }\textbf {\bibinfo {volume} {4}},\ \bibinfo {pages} {1--10}
  (\bibinfo {year} {2018})}\BibitemShut {NoStop}%
\bibitem [{ \citenamefont {Hormozi}\ \emph {et~al.}(2017) \citenamefont
  {Hormozi},  \citenamefont {Brown},  \citenamefont {Carleo},\ and\  \citenamefont
  {Troyer}}]{nonstoquastic}%
  \BibitemOpen
  \bibfield  {author} {\bibinfo {author} {\bibfnamefont {Layla}\ \bibnamefont
  {Hormozi}}, \bibinfo {author} {\bibfnamefont {Ethan~W}\ \bibnamefont
  {Brown}}, \bibinfo {author} {\bibfnamefont {Giuseppe}\ \bibnamefont
  {Carleo}}, \ and\ \bibinfo {author} {\bibfnamefont {Matthias}\ \bibnamefont
  {Troyer}},\ }\bibfield  {title} {\enquote {\bibinfo {title} {Nonstoquastic
  hamiltonians and quantum annealing of an ising spin glass},}\ }\href@noop {}
  {\bibfield  {journal} {\bibinfo  {journal} {Phys. Rev. B}\ }\textbf {\bibinfo
  {volume} {95}},\ \bibinfo {pages} {184416} (\bibinfo {year}
  {2017})}\BibitemShut {NoStop}%
\bibitem [{ \citenamefont {Herr}\ \emph {et~al.}(2017) \citenamefont {Herr},
   \citenamefont {Brown},  \citenamefont {Heim},  \citenamefont {Konz},
   \citenamefont {Mazzola},\ and\  \citenamefont
  {Troyer}}]{Herr2017OptimizingSF}%
  \BibitemOpen
  \bibfield  {author} {\bibinfo {author} {\bibfnamefont {Daniel}\ \bibnamefont
  {Herr}}, \bibinfo {author} {\bibfnamefont {Ethan}\ \bibnamefont {Brown}},
  \bibinfo {author} {\bibfnamefont {Bettina}\ \bibnamefont {Heim}}, \bibinfo
  {author} {\bibfnamefont {M.}~\bibnamefont {Konz}}, \bibinfo {author}
  {\bibfnamefont {Guglielmo}\ \bibnamefont {Mazzola}}, \ and\ \bibinfo {author}
  {\bibfnamefont {Matthias}\ \bibnamefont {Troyer}},\ }\href@noop {} {\enquote
  {\bibinfo {title} {Optimizing schedules for quantum annealing},}\ } (\bibinfo
  {year} {2017}),\ \Eprint {http://arxiv.org/abs/arXiv:1312.5602}
  {arXiv:1312.5602} \BibitemShut {NoStop}%
\bibitem [{ \citenamefont {Zeng}\ \emph {et~al.}(2016) \citenamefont {Zeng},
   \citenamefont {Zhang},\ and\  \citenamefont {Sarovar}}]{zeng2016schedule}%
  \BibitemOpen
  \bibfield  {author} {\bibinfo {author} {\bibfnamefont {Lishan}\ \bibnamefont
  {Zeng}}, \bibinfo {author} {\bibfnamefont {Jun}\ \bibnamefont {Zhang}}, \
  and\ \bibinfo {author} {\bibfnamefont {Mohan}\ \bibnamefont {Sarovar}},\
  }\bibfield  {title} {\enquote {\bibinfo {title} {Schedule path optimization
  for adiabatic quantum computing and optimization},}\ }\href@noop {}
  {\bibfield  {journal} {\bibinfo  {journal} {J Phys. A: Math Theor.}\ }\textbf
  {\bibinfo {volume} {49}},\ \bibinfo {pages} {165305} (\bibinfo {year}
  {2016})}\BibitemShut {NoStop}%
\bibitem [{ \citenamefont {Susa}\ \emph {et~al.}(2018) \citenamefont {Susa},
   \citenamefont {Yamashiro},  \citenamefont {Yamamoto},\ and\  \citenamefont
  {Nishimori}}]{susa2018exponential}%
  \BibitemOpen
  \bibfield  {author} {\bibinfo {author} {\bibfnamefont {Yuki}\ \bibnamefont
  {Susa}}, \bibinfo {author} {\bibfnamefont {Yu}~\bibnamefont {Yamashiro}},
  \bibinfo {author} {\bibfnamefont {Masayuki}\ \bibnamefont {Yamamoto}}, \ and\
  \bibinfo {author} {\bibfnamefont {Hidetoshi}\ \bibnamefont {Nishimori}},\
  }\bibfield  {title} {\enquote {\bibinfo {title} {Exponential speedup of
  quantum annealing by inhomogeneous driving of the transverse field},}\
  }\href@noop {} {\bibfield  {journal} {\bibinfo  {journal} {J. Phys. Sos.
  Jpn.}\ }\textbf {\bibinfo {volume} {87}},\ \bibinfo {pages} {023002}
  (\bibinfo {year} {2018})}\BibitemShut {NoStop}%
\bibitem [{ \citenamefont {Albash}\ and\  \citenamefont
  {Lidar}(2018)}]{Albash2018}%
  \BibitemOpen
  \bibfield  {author} {\bibinfo {author} {\bibfnamefont {Tameem}\ \bibnamefont
  {Albash}}\ and\ \bibinfo {author} {\bibfnamefont {Daniel~A}\ \bibnamefont
  {Lidar}},\ }\bibfield  {title} {\enquote {\bibinfo {title} {Adiabatic quantum
  computation},}\ }\href@noop {} {\bibfield  {journal} {\bibinfo  {journal}
  {Rev. Mod. Phys.}\ }\textbf {\bibinfo {volume} {90}},\ \bibinfo {pages}
  {015002} (\bibinfo {year} {2018})}\BibitemShut {NoStop}%
\bibitem [{ \citenamefont {Hauke}\ \emph {et~al.}(2019) \citenamefont {Hauke},
   \citenamefont {Katzgraber},  \citenamefont {Lechner},  \citenamefont
  {Nishimori},\ and\  \citenamefont {Oliver}}]{hauke2019perspectives}%
  \BibitemOpen
  \bibfield  {author} {\bibinfo {author} {\bibfnamefont {Philipp}\ \bibnamefont
  {Hauke}}, \bibinfo {author} {\bibfnamefont {Helmut~G}\ \bibnamefont
  {Katzgraber}}, \bibinfo {author} {\bibfnamefont {Wolfgang}\ \bibnamefont
  {Lechner}}, \bibinfo {author} {\bibfnamefont {Hidetoshi}\ \bibnamefont
  {Nishimori}}, \ and\ \bibinfo {author} {\bibfnamefont {William~D}\
  \bibnamefont {Oliver}},\ }\bibfield  {title} {\enquote {\bibinfo {title}
  {Perspectives of quantum annealing: Methods and implementations},}\
  }\href@noop {} {\  (\bibinfo {year} {2019})},\ \Eprint
  {http://arxiv.org/abs/arXiv:1903.06559} {arXiv:1903.06559} \BibitemShut
  {NoStop}%
 \bibitem{dalgaard2020global}
M.~Dalgaard, F.~Motzoi, J.~J. S{\o}rensen, and J.~Sherson, ``Global
  optimization of quantum dynamics with alphazero deep exploration,'' \emph{npj
  Quantum Information}, vol.~6, no.~1, pp. 1--9, 2020.

\bibitem{susa2021variational}
Y.~Susa and H.~Nishimori, ``Variational optimization of the quantum annealing
  schedule for the lechner-hauke-zoller scheme,'' \emph{Physical Review A},
  vol. 103, no.~2, p. 022619, 2021.

\bibitem{herr2017optimizing}
D.~Herr, E.~Brown, B.~Heim, M.~K{\"o}nz, G.~Mazzola, and M.~Troyer,
  ``Optimizing schedules for quantum annealing,'' \emph{arXiv preprint
  arXiv:1705.00420}, 2017.

\bibitem{schiffer2021adiabatic}
B.~F. Schiffer, J.~Tura, and J.~I. Cirac, ``Adiabatic spectroscopy and a
  variational quantum adiabatic algorithm,'' \emph{arXiv preprint
  arXiv:2103.01226}, 2021.

\bibitem{boixo2009eigenpath}
S.~Boixo, E.~Knill, R.~D. Somma \emph{et~al.}, ``Eigenpath traversal by phase
  randomization.'' \emph{Quantum Inf. Comput.}, vol.~9, no. 9\&10, pp.
  833--855, 2009.

\bibitem{caneva2011chopped}
T.~Caneva, T.~Calarco, and S.~Montangero, ``Chopped random-basis quantum
  optimization,'' \emph{Physical Review A}, vol.~84, no.~2, p. 022326, 2011.
  
\bibitem [{ \citenamefont {Farhi}\ \emph {et~al.}(2000) \citenamefont {Farhi},
   \citenamefont {Goldstone},  \citenamefont {Gutmann},\ and\  \citenamefont
  {Sipser}}]{quant-ph/0001106}%
  \BibitemOpen
  \bibfield  {author} {\bibinfo {author} {\bibfnamefont {Edward}\ \bibnamefont
  {Farhi}}, \bibinfo {author} {\bibfnamefont {Jeffrey}\ \bibnamefont
  {Goldstone}}, \bibinfo {author} {\bibfnamefont {Sam}\ \bibnamefont
  {Gutmann}}, \ and\ \bibinfo {author} {\bibfnamefont {Michael}\ \bibnamefont
  {Sipser}},\ }\href@noop {} {\enquote {\bibinfo {title} {Quantum computation
  by adiabatic evolution},}\ } (\bibinfo {year} {2000}),\ \Eprint
  {http://arxiv.org/abs/arXiv:quant-ph/0001106} {arXiv:quant-ph/0001106}
  \BibitemShut {NoStop}%
\bibitem [{ \citenamefont {Farhi}\ \emph {et~al.}(2001) \citenamefont {Farhi},
   \citenamefont {Goldstone},  \citenamefont {Gutmann},  \citenamefont {Lapan},
   \citenamefont {Lundgren},\ and\  \citenamefont {Preda}}]{Farhi2001}%
  \BibitemOpen
  \bibfield  {author} {\bibinfo {author} {\bibfnamefont {E.}~\bibnamefont
  {Farhi}}, \bibinfo {author} {\bibfnamefont {J.}~\bibnamefont {Goldstone}},
  \bibinfo {author} {\bibfnamefont {S.}~\bibnamefont {Gutmann}}, \bibinfo
  {author} {\bibfnamefont {J.}~\bibnamefont {Lapan}}, \bibinfo {author}
  {\bibfnamefont {A.}~\bibnamefont {Lundgren}}, \ and\ \bibinfo {author}
  {\bibfnamefont {D.}~\bibnamefont {Preda}},\ }\bibfield  {title} {\enquote
  {\bibinfo {title} {A quantum adiabatic evolution algorithm applied to random
  instances of an {NP}-complete problem},}\ }\href {\doibase
  10.1126/science.1057726} {\bibfield  {journal} {\bibinfo  {journal}
  {Science}\ }\textbf {\bibinfo {volume} {292}},\ \bibinfo {pages} {472--475}
  (\bibinfo {year} {2001})}\BibitemShut {NoStop}%
\bibitem [{ \citenamefont {Das}\ and\  \citenamefont
  {Chakrabarti}(2005)}]{article}%
  \BibitemOpen
  \bibfield  {author} {\bibinfo {author} {\bibfnamefont {Arnab}\ \bibnamefont
  {Das}}\ and\ \bibinfo {author} {\bibfnamefont {BK}~\bibnamefont
  {Chakrabarti}},\ }\href@noop {} {\emph {\bibinfo {title} {Quantum annealing
  and related optimization methods}}}\ (\bibinfo  {publisher} {Springer},\
  \bibinfo {year} {2005})\BibitemShut {NoStop}%
\bibitem [{ \citenamefont {Childs}\ \emph {et~al.}(2001) \citenamefont {Childs},
   \citenamefont {Farhi},\ and\  \citenamefont {Preskill}}]{Childs2001}%
  \BibitemOpen
  \bibfield  {author} {\bibinfo {author} {\bibfnamefont {Andrew~M}\
  \bibnamefont {Childs}}, \bibinfo {author} {\bibfnamefont {Edward}\
  \bibnamefont {Farhi}}, \ and\ \bibinfo {author} {\bibfnamefont {John}\
  \bibnamefont {Preskill}},\ }\bibfield  {title} {\enquote {\bibinfo {title}
  {Robustness of adiabatic quantum computation},}\ }\href@noop {} {\bibfield
  {journal} {\bibinfo  {journal} {Phys. Rev. A}\ }\textbf {\bibinfo {volume}
  {65}},\ \bibinfo {pages} {012322} (\bibinfo {year} {2001})}\BibitemShut
  {NoStop}%
\bibitem [{ \citenamefont {Aharonov}\ \emph {et~al.}(2004) \citenamefont
  {Aharonov},  \citenamefont {van Dam},  \citenamefont {Kempe},  \citenamefont
  {Landau},  \citenamefont {Lloyd},\ and\  \citenamefont
  {Regev}}]{quant-ph/0405098}%
  \BibitemOpen
  \bibfield  {author} {\bibinfo {author} {\bibfnamefont {Dorit}\ \bibnamefont
  {Aharonov}}, \bibinfo {author} {\bibfnamefont {Wim}\ \bibnamefont {van Dam}},
  \bibinfo {author} {\bibfnamefont {Julia}\ \bibnamefont {Kempe}}, \bibinfo
  {author} {\bibfnamefont {Zeph}\ \bibnamefont {Landau}}, \bibinfo {author}
  {\bibfnamefont {Seth}\ \bibnamefont {Lloyd}}, \ and\ \bibinfo {author}
  {\bibfnamefont {Oded}\ \bibnamefont {Regev}},\ }\bibfield  {title} {\enquote
  {\bibinfo {title} {Adiabatic quantum computation is equivalent to standard
  quantum computation},}\ }\href@noop {} {\  (\bibinfo {year} {2004})},\
  \Eprint {http://arxiv.org/abs/arXiv:quant-ph/0405098}
  {arXiv:quant-ph/0405098} \BibitemShut {NoStop}%
\bibitem [{ \citenamefont {Coulom}(2007)}]{Coulom2007}%
  \BibitemOpen
  \bibfield  {author} {\bibinfo {author} {\bibfnamefont {R{\'{e}}mi}\
  \bibnamefont {Coulom}},\ }\bibfield  {title} {\enquote {\bibinfo {title}
  {Efficient selectivity and backup operators in monte-carlo tree search},}\
  }in\ \href {\doibase 10.1007/978-3-540-75538-8_7} {\emph {\bibinfo
  {booktitle} {Computers and Games}}}\ (\bibinfo  {publisher} {Springer Berlin
  Heidelberg},\ \bibinfo {year} {2007})\ pp.\ \bibinfo {pages}
  {72--83}\BibitemShut {NoStop}%
\bibitem [{ \citenamefont {Kocsis}\ and\  \citenamefont
  {Szepesv{\'a}ri}(2006)}]{10.1007/11871842_29}%
  \BibitemOpen
  \bibfield  {author} {\bibinfo {author} {\bibfnamefont {Levente}\ \bibnamefont
  {Kocsis}}\ and\ \bibinfo {author} {\bibfnamefont {Csaba}\ \bibnamefont
  {Szepesv{\'a}ri}},\ }\bibfield  {title} {\enquote {\bibinfo {title} {Bandit
  based monte-carlo planning},}\ }in\ \href@noop {} {\emph {\bibinfo
  {booktitle} {European conference on machine learning}}}\ (\bibinfo
  {organization} {Springer},\ \bibinfo {address} {Berlin, Heidelberg},\
  \bibinfo {year} {2006})\ pp.\ \bibinfo {pages} {282--293}\BibitemShut
  {NoStop}%
\bibitem [{ \citenamefont {Kocsis}\ and\  \citenamefont
  {Szepesv{\'{a}}ri}(2006)}]{Kocsis2006}%
  \BibitemOpen
  \bibfield  {author} {\bibinfo {author} {\bibfnamefont {Levente}\ \bibnamefont
  {Kocsis}}\ and\ \bibinfo {author} {\bibfnamefont {Csaba}\ \bibnamefont
  {Szepesv{\'{a}}ri}},\ }\bibfield  {title} {\enquote {\bibinfo {title} {Bandit
  based monte-carlo planning},}\ }in\ \href {\doibase 10.1007/11871842_29}
  {\emph {\bibinfo {booktitle} {Lecture Notes in Computer Science}}}\ (\bibinfo
   {publisher} {Springer Berlin Heidelberg},\ \bibinfo {year} {2006})\ pp.\
  \bibinfo {pages} {282--293}\BibitemShut {NoStop}%
\bibitem [{ \citenamefont {Lee}\ \emph {et~al.}(2009) \citenamefont {Lee},
   \citenamefont {Wang},  \citenamefont {Chaslot},  \citenamefont {Hoock},
   \citenamefont {Rimmel},  \citenamefont {Teytaud},  \citenamefont {Tsai},
   \citenamefont {Hsu},\ and\  \citenamefont {Hong}}]{ChangShingLee2009}%
  \BibitemOpen
  \bibfield  {author} {\bibinfo {author} {\bibfnamefont {Chang-Shing}\
  \bibnamefont {Lee}}, \bibinfo {author} {\bibfnamefont {Mei-Hui}\ \bibnamefont
  {Wang}}, \bibinfo {author} {\bibfnamefont {Guillaume}\ \bibnamefont
  {Chaslot}}, \bibinfo {author} {\bibfnamefont {Jean-Baptiste}\ \bibnamefont
  {Hoock}}, \bibinfo {author} {\bibfnamefont {Arpad}\ \bibnamefont {Rimmel}},
  \bibinfo {author} {\bibfnamefont {Olivier}\ \bibnamefont {Teytaud}}, \bibinfo
  {author} {\bibfnamefont {Shang-Rong}\ \bibnamefont {Tsai}}, \bibinfo {author}
  {\bibfnamefont {Shun-Chin}\ \bibnamefont {Hsu}}, \ and\ \bibinfo {author}
  {\bibfnamefont {Tzung-Pei}\ \bibnamefont {Hong}},\ }\bibfield  {title}
  {\enquote {\bibinfo {title} {The computational intelligence of mogo revealed
  in taiwan's computer go tournaments},}\ }\href@noop {} {\bibfield  {journal}
  {\bibinfo  {journal} {IEEE Transactions on Computational Intelligence and AI
  in games}\ }\textbf {\bibinfo {volume} {1}},\ \bibinfo {pages} {73--89}
  (\bibinfo {year} {2009})}\BibitemShut {NoStop}%
\bibitem [{ \citenamefont {Silver}\ \emph {et~al.}(2017) \citenamefont {Silver},
   \citenamefont {Schrittwieser},  \citenamefont {Simonyan},  \citenamefont
  {Antonoglou},  \citenamefont {Huang},  \citenamefont {Guez},  \citenamefont
  {Hubert},  \citenamefont {Baker},  \citenamefont {Lai},  \citenamefont {Bolton},
   \citenamefont {Chen},  \citenamefont {Lillicrap},  \citenamefont {Hui},
   \citenamefont {Sifre},  \citenamefont {van~den Driessche},  \citenamefont
  {Graepel},\ and\  \citenamefont {Hassabis}}]{Silver2017}%
  \BibitemOpen
  \bibfield  {author} {\bibinfo {author} {\bibfnamefont {David}\ \bibnamefont
  {Silver}}, \bibinfo {author} {\bibfnamefont {Julian}\ \bibnamefont
  {Schrittwieser}}, \bibinfo {author} {\bibfnamefont {Karen}\ \bibnamefont
  {Simonyan}}, \bibinfo {author} {\bibfnamefont {Ioannis}\ \bibnamefont
  {Antonoglou}}, \bibinfo {author} {\bibfnamefont {Aja}\ \bibnamefont {Huang}},
  \bibinfo {author} {\bibfnamefont {Arthur}\ \bibnamefont {Guez}}, \bibinfo
  {author} {\bibfnamefont {Thomas}\ \bibnamefont {Hubert}}, \bibinfo {author}
  {\bibfnamefont {Lucas}\ \bibnamefont {Baker}}, \bibinfo {author}
  {\bibfnamefont {Matthew}\ \bibnamefont {Lai}}, \bibinfo {author}
  {\bibfnamefont {Adrian}\ \bibnamefont {Bolton}}, \bibinfo {author}
  {\bibfnamefont {Yutian}\ \bibnamefont {Chen}}, \bibinfo {author}
  {\bibfnamefont {Timothy}\ \bibnamefont {Lillicrap}}, \bibinfo {author}
  {\bibfnamefont {Fan}\ \bibnamefont {Hui}}, \bibinfo {author} {\bibfnamefont
  {Laurent}\ \bibnamefont {Sifre}}, \bibinfo {author} {\bibfnamefont {George}\
  \bibnamefont {van~den Driessche}}, \bibinfo {author} {\bibfnamefont {Thore}\
  \bibnamefont {Graepel}}, \ and\ \bibinfo {author} {\bibfnamefont {Demis}\
  \bibnamefont {Hassabis}},\ }\bibfield  {title} {\enquote {\bibinfo {title}
  {Mastering the game of go without human knowledge},}\ }\href {\doibase
  10.1038/nature24270} {\bibfield  {journal} {\bibinfo  {journal} {Nature}\
  }\textbf {\bibinfo {volume} {550}},\ \bibinfo {pages} {354--359} (\bibinfo
  {year} {2017})}\BibitemShut {NoStop}%
\bibitem [{ \citenamefont {Silver}\ \emph {et~al.}(2018) \citenamefont {Silver},
   \citenamefont {Hubert},  \citenamefont {Schrittwieser},  \citenamefont
  {Antonoglou},  \citenamefont {Lai},  \citenamefont {Guez},  \citenamefont
  {Lanctot},  \citenamefont {Sifre},  \citenamefont {Kumaran},  \citenamefont
  {Graepel},  \citenamefont {Lillicrap},  \citenamefont {Simonyan},\ and\
   \citenamefont {Hassabis}}]{Silver2018}%
  \BibitemOpen
  \bibfield  {author} {\bibinfo {author} {\bibfnamefont {David}\ \bibnamefont
  {Silver}}, \bibinfo {author} {\bibfnamefont {Thomas}\ \bibnamefont {Hubert}},
  \bibinfo {author} {\bibfnamefont {Julian}\ \bibnamefont {Schrittwieser}},
  \bibinfo {author} {\bibfnamefont {Ioannis}\ \bibnamefont {Antonoglou}},
  \bibinfo {author} {\bibfnamefont {Matthew}\ \bibnamefont {Lai}}, \bibinfo
  {author} {\bibfnamefont {Arthur}\ \bibnamefont {Guez}}, \bibinfo {author}
  {\bibfnamefont {Marc}\ \bibnamefont {Lanctot}}, \bibinfo {author}
  {\bibfnamefont {Laurent}\ \bibnamefont {Sifre}}, \bibinfo {author}
  {\bibfnamefont {Dharshan}\ \bibnamefont {Kumaran}}, \bibinfo {author}
  {\bibfnamefont {Thore}\ \bibnamefont {Graepel}}, \bibinfo {author}
  {\bibfnamefont {Timothy}\ \bibnamefont {Lillicrap}}, \bibinfo {author}
  {\bibfnamefont {Karen}\ \bibnamefont {Simonyan}}, \ and\ \bibinfo {author}
  {\bibfnamefont {Demis}\ \bibnamefont {Hassabis}},\ }\bibfield  {title}
  {\enquote {\bibinfo {title} {A general reinforcement learning algorithm that
  masters chess, shogi, and go through self-play},}\ }\href {\doibase
  10.1126/science.aar6404} {\bibfield  {journal} {\bibinfo  {journal}
  {Science}\ }\textbf {\bibinfo {volume} {362}},\ \bibinfo {pages} {1140--1144}
  (\bibinfo {year} {2018})}\BibitemShut {NoStop}%
\bibitem [{ \citenamefont {Peruzzo}\ \emph {et~al.}(2014) \citenamefont
  {Peruzzo},  \citenamefont {McClean},  \citenamefont {Shadbolt},  \citenamefont
  {Yung},  \citenamefont {Zhou},  \citenamefont {Love},  \citenamefont
  {Aspuru-Guzik},\ and\  \citenamefont {O’brien}}]{Peruzzo2014}%
  \BibitemOpen
  \bibfield  {author} {\bibinfo {author} {\bibfnamefont {Alberto}\ \bibnamefont
  {Peruzzo}}, \bibinfo {author} {\bibfnamefont {Jarrod}\ \bibnamefont
  {McClean}}, \bibinfo {author} {\bibfnamefont {Peter}\ \bibnamefont
  {Shadbolt}}, \bibinfo {author} {\bibfnamefont {Man-Hong}\ \bibnamefont
  {Yung}}, \bibinfo {author} {\bibfnamefont {Xiao-Qi}\ \bibnamefont {Zhou}},
  \bibinfo {author} {\bibfnamefont {Peter~J}\ \bibnamefont {Love}}, \bibinfo
  {author} {\bibfnamefont {Al{\'a}n}\ \bibnamefont {Aspuru-Guzik}}, \ and\
  \bibinfo {author} {\bibfnamefont {Jeremy~L}\ \bibnamefont {O’brien}},\
  }\bibfield  {title} {\enquote {\bibinfo {title} {A variational eigenvalue
  solver on a photonic quantum processor},}\ }\href@noop {} {\bibfield
  {journal} {\bibinfo  {journal} {Nat. Commun.}\ }\textbf {\bibinfo {volume}
  {5}},\ \bibinfo {pages} {4213} (\bibinfo {year} {2014})}\BibitemShut
  {NoStop}%
\bibitem [{ \citenamefont {Kandala}\ \emph {et~al.}(2017) \citenamefont
  {Kandala},  \citenamefont {Mezzacapo},  \citenamefont {Temme},  \citenamefont
  {Takita},  \citenamefont {Brink},  \citenamefont {Chow},\ and\  \citenamefont
  {Gambetta}}]{Kandala2017}%
  \BibitemOpen
  \bibfield  {author} {\bibinfo {author} {\bibfnamefont {Abhinav}\ \bibnamefont
  {Kandala}}, \bibinfo {author} {\bibfnamefont {Antonio}\ \bibnamefont
  {Mezzacapo}}, \bibinfo {author} {\bibfnamefont {Kristan}\ \bibnamefont
  {Temme}}, \bibinfo {author} {\bibfnamefont {Maika}\ \bibnamefont {Takita}},
  \bibinfo {author} {\bibfnamefont {Markus}\ \bibnamefont {Brink}}, \bibinfo
  {author} {\bibfnamefont {Jerry~M.}\ \bibnamefont {Chow}}, \ and\ \bibinfo
  {author} {\bibfnamefont {Jay~M.}\ \bibnamefont {Gambetta}},\ }\bibfield
  {title} {\enquote {\bibinfo {title} {Hardware-efficient variational quantum
  eigensolver for small molecules and quantum magnets},}\ }\href {\doibase
  10.1038/nature23879} {\bibfield  {journal} {\bibinfo  {journal} {Nature}\
  }\textbf {\bibinfo {volume} {549}},\ \bibinfo {pages} {242--246} (\bibinfo
  {year} {2017})}\BibitemShut {NoStop}%
\bibitem [{ \citenamefont {Farhi}\ \emph {et~al.}(2014) \citenamefont {Farhi},
   \citenamefont {Goldstone},\ and\  \citenamefont {Gutmann}}]{1411.4028}%
  \BibitemOpen
  \bibfield  {author} {\bibinfo {author} {\bibfnamefont {Edward}\ \bibnamefont
  {Farhi}}, \bibinfo {author} {\bibfnamefont {Jeffrey}\ \bibnamefont
  {Goldstone}}, \ and\ \bibinfo {author} {\bibfnamefont {Sam}\ \bibnamefont
  {Gutmann}},\ }\href@noop {} {\enquote {\bibinfo {title} {A quantum
  approximate optimization algorithm},}\ } (\bibinfo {year} {2014}),\ \Eprint
  {http://arxiv.org/abs/arXiv:1411.4028} {arXiv:1411.4028} \BibitemShut
  {NoStop}%
\bibitem [{ \citenamefont {McClean}\ \emph {et~al.}(2016) \citenamefont
  {McClean},  \citenamefont {Romero},  \citenamefont {Babbush},\ and\
   \citenamefont {Aspuru-Guzik}}]{McClean2016}%
  \BibitemOpen
  \bibfield  {author} {\bibinfo {author} {\bibfnamefont {Jarrod~R}\
  \bibnamefont {McClean}}, \bibinfo {author} {\bibfnamefont {Jonathan}\
  \bibnamefont {Romero}}, \bibinfo {author} {\bibfnamefont {Ryan}\ \bibnamefont
  {Babbush}}, \ and\ \bibinfo {author} {\bibfnamefont {Al{\'{a}}n}\
  \bibnamefont {Aspuru-Guzik}},\ }\bibfield  {title} {\enquote {\bibinfo
  {title} {The theory of variational hybrid quantum-classical algorithms},}\
  }\href {\doibase 10.1088/1367-2630/18/2/023023} {\bibfield  {journal}
  {\bibinfo  {journal} {New. J. Phys.}\ }\textbf {\bibinfo {volume} {18}},\
  \bibinfo {pages} {023023} (\bibinfo {year} {2016})}\BibitemShut {NoStop}%
\bibitem [{ \citenamefont {Preskill}(2018)}]{Preskill2018}%
  \BibitemOpen
  \bibfield  {author} {\bibinfo {author} {\bibfnamefont {John}\ \bibnamefont
  {Preskill}},\ }\bibfield  {title} {\enquote {\bibinfo {title} {Quantum
  computing in the {NISQ} era and beyond},}\ }\href {\doibase
  10.22331/q-2018-08-06-79} {\bibfield  {journal} {\bibinfo  {journal}
  {Quantum}\ }\textbf {\bibinfo {volume} {2}},\ \bibinfo {pages} {79} (\bibinfo
  {year} {2018})}\BibitemShut {NoStop}%
\bibitem [{ \citenamefont {Cao}\ \emph {et~al.}(2019) \citenamefont {Cao},
   \citenamefont {Romero},  \citenamefont {Olson},  \citenamefont {Degroote},
   \citenamefont {Johnson},  \citenamefont {Kieferov{\'{a}}},  \citenamefont
  {Kivlichan},  \citenamefont {Menke},  \citenamefont {Peropadre},  \citenamefont
  {Sawaya},  \citenamefont {Sim},  \citenamefont {Veis},\ and\  \citenamefont
  {Aspuru-Guzik}}]{Cao2019}%
  \BibitemOpen
  \bibfield  {author} {\bibinfo {author} {\bibfnamefont {Yudong}\ \bibnamefont
  {Cao}}, \bibinfo {author} {\bibfnamefont {Jonathan}\ \bibnamefont {Romero}},
  \bibinfo {author} {\bibfnamefont {Jonathan~P.}\ \bibnamefont {Olson}},
  \bibinfo {author} {\bibfnamefont {Matthias}\ \bibnamefont {Degroote}},
  \bibinfo {author} {\bibfnamefont {Peter~D.}\ \bibnamefont {Johnson}},
  \bibinfo {author} {\bibfnamefont {M{\'{a}}ria}\ \bibnamefont
  {Kieferov{\'{a}}}}, \bibinfo {author} {\bibfnamefont {Ian~D.}\ \bibnamefont
  {Kivlichan}}, \bibinfo {author} {\bibfnamefont {Tim}\ \bibnamefont {Menke}},
  \bibinfo {author} {\bibfnamefont {Borja}\ \bibnamefont {Peropadre}}, \bibinfo
  {author} {\bibfnamefont {Nicolas P.~D.}\ \bibnamefont {Sawaya}}, \bibinfo
  {author} {\bibfnamefont {Sukin}\ \bibnamefont {Sim}}, \bibinfo {author}
  {\bibfnamefont {Libor}\ \bibnamefont {Veis}}, \ and\ \bibinfo {author}
  {\bibfnamefont {Al{\'{a}}n}\ \bibnamefont {Aspuru-Guzik}},\ }\bibfield
  {title} {\enquote {\bibinfo {title} {Quantum chemistry in the age of quantum
  computing},}\ }\href {\doibase 10.1021/acs.chemrev.8b00803} {\bibfield
  {journal} {\bibinfo  {journal} {Chem. Rev.}\ }\textbf {\bibinfo {volume}
  {119}},\ \bibinfo {pages} {10856--10915} (\bibinfo {year}
  {2019})}\BibitemShut {NoStop}%
\bibitem [{ \citenamefont {Chen}\ \emph {et~al.}(2019) \citenamefont {Chen},
   \citenamefont {Gong},  \citenamefont {Xu},  \citenamefont {Yuan},  \citenamefont
  {Wang},  \citenamefont {Wang},  \citenamefont {Ying},  \citenamefont {Lin},
   \citenamefont {Xu},  \citenamefont {Wu} \emph {et~al.}}]{1905.03150}%
  \BibitemOpen
  \bibfield  {author} {\bibinfo {author} {\bibfnamefont {Ming-Cheng}\
  \bibnamefont {Chen}}, \bibinfo {author} {\bibfnamefont {Ming}\ \bibnamefont
  {Gong}}, \bibinfo {author} {\bibfnamefont {Xiao-Si}\ \bibnamefont {Xu}},
  \bibinfo {author} {\bibfnamefont {Xiao}\ \bibnamefont {Yuan}}, \bibinfo
  {author} {\bibfnamefont {Jian-Wen}\ \bibnamefont {Wang}}, \bibinfo {author}
  {\bibfnamefont {Can}\ \bibnamefont {Wang}}, \bibinfo {author} {\bibfnamefont
  {Chong}\ \bibnamefont {Ying}}, \bibinfo {author} {\bibfnamefont {Jin}\
  \bibnamefont {Lin}}, \bibinfo {author} {\bibfnamefont {Yu}~\bibnamefont
  {Xu}}, \bibinfo {author} {\bibfnamefont {Yulin}\ \bibnamefont {Wu}},  \emph
  {et~al.},\ }\bibfield  {title} {\enquote {\bibinfo {title} {Demonstration of
  adiabatic variational quantum computing with a superconducting quantum
  coprocessor},}\ }\href@noop {} {\bibfield  {journal} {\bibinfo  {journal}
  {arXiv preprint arXiv:1905.03150}\ } (\bibinfo {year} {2019})}\BibitemShut
  {NoStop}%
\bibitem [{ \citenamefont {Zhang}\ \emph {et~al.}(2019) \citenamefont {Zhang},
   \citenamefont {Wei},  \citenamefont {Asad},  \citenamefont {Yang},\ and\
   \citenamefont {Wang}}]{zhang2019reinforcement}%
  \BibitemOpen
  \bibfield  {author} {\bibinfo {author} {\bibfnamefont {Xiao-Ming}\
  \bibnamefont {Zhang}}, \bibinfo {author} {\bibfnamefont {Zezhu}\ \bibnamefont
  {Wei}}, \bibinfo {author} {\bibfnamefont {Raza}\ \bibnamefont {Asad}},
  \bibinfo {author} {\bibfnamefont {Xu-Chen}\ \bibnamefont {Yang}}, \ and\
  \bibinfo {author} {\bibfnamefont {Xin}\ \bibnamefont {Wang}},\ }\bibfield
  {title} {\enquote {\bibinfo {title} {When reinforcement learning stands out
  in quantum control? a comparative study on state preparation},}\ }\href@noop
  {} {\bibfield  {journal} {\bibinfo  {journal} {arXiv preprint
  arXiv:1902.02157}\ } (\bibinfo {year} {2019})}\BibitemShut {NoStop}%
\bibitem [{ \citenamefont {Chen}\ \emph {et~al.}(2014) \citenamefont {Chen},
   \citenamefont {Dong},  \citenamefont {Li},  \citenamefont {Chu},\ and\
   \citenamefont {Tarn}}]{ChunlinChen2014}%
  \BibitemOpen
  \bibfield  {author} {\bibinfo {author} {\bibfnamefont {Chunlin}\ \bibnamefont
  {Chen}}, \bibinfo {author} {\bibfnamefont {Daoyi}\ \bibnamefont {Dong}},
  \bibinfo {author} {\bibfnamefont {Han-Xiong}\ \bibnamefont {Li}}, \bibinfo
  {author} {\bibfnamefont {Jian}\ \bibnamefont {Chu}}, \ and\ \bibinfo {author}
  {\bibfnamefont {Tzyh-Jong}\ \bibnamefont {Tarn}},\ }\bibfield  {title}
  {\enquote {\bibinfo {title} {Fidelity-based probabilistic q-learning for
  control of quantum systems},}\ }\href {\doibase 10.1109/tnnls.2013.2283574}
  {\bibfield  {journal} {\bibinfo  {journal} {{IEEE} Transactions on Neural
  Networks and Learning Systems}\ }\textbf {\bibinfo {volume} {25}},\ \bibinfo
  {pages} {920--933} (\bibinfo {year} {2014})}\BibitemShut {NoStop}%
\bibitem [{ \citenamefont {Bukov}\ \emph {et~al.}(2018) \citenamefont {Bukov},
   \citenamefont {Day},  \citenamefont {Sels},  \citenamefont {Weinberg},
   \citenamefont {Polkovnikov},\ and\  \citenamefont {Mehta}}]{Bukov2018}%
  \BibitemOpen
  \bibfield  {author} {\bibinfo {author} {\bibfnamefont {Marin}\ \bibnamefont
  {Bukov}}, \bibinfo {author} {\bibfnamefont {Alexandre~GR}\ \bibnamefont
  {Day}}, \bibinfo {author} {\bibfnamefont {Dries}\ \bibnamefont {Sels}},
  \bibinfo {author} {\bibfnamefont {Phillip}\ \bibnamefont {Weinberg}},
  \bibinfo {author} {\bibfnamefont {Anatoli}\ \bibnamefont {Polkovnikov}}, \
  and\ \bibinfo {author} {\bibfnamefont {Pankaj}\ \bibnamefont {Mehta}},\
  }\bibfield  {title} {\enquote {\bibinfo {title} {Reinforcement learning in
  different phases of quantum control},}\ }\href@noop {} {\bibfield  {journal}
  {\bibinfo  {journal} {Phys. Rev. X}\ }\textbf {\bibinfo {volume} {8}},\
  \bibinfo {pages} {031086} (\bibinfo {year} {2018})}\BibitemShut {NoStop}%
\bibitem [{ \citenamefont {Niu}\ \emph {et~al.}(2019) \citenamefont {Niu},
   \citenamefont {Boixo},  \citenamefont {Smelyanskiy},\ and\  \citenamefont
  {Neven}}]{Niu2019}%
  \BibitemOpen
  \bibfield  {author} {\bibinfo {author} {\bibfnamefont {Murphy~Yuezhen}\
  \bibnamefont {Niu}}, \bibinfo {author} {\bibfnamefont {Sergio}\ \bibnamefont
  {Boixo}}, \bibinfo {author} {\bibfnamefont {Vadim~N}\ \bibnamefont
  {Smelyanskiy}}, \ and\ \bibinfo {author} {\bibfnamefont {Hartmut}\
  \bibnamefont {Neven}},\ }\bibfield  {title} {\enquote {\bibinfo {title}
  {Universal quantum control through deep reinforcement learning},}\
  }\href@noop {} {\bibfield  {journal} {\bibinfo  {journal} {npj Quantum Inf.}\
  }\textbf {\bibinfo {volume} {5}},\ \bibinfo {pages} {1--8} (\bibinfo {year}
  {2019})}\BibitemShut {NoStop}%
\bibitem [{ \citenamefont {McKiernan}\ \emph {et~al.}(2019) \citenamefont
  {McKiernan},  \citenamefont {Davis},  \citenamefont {Alam},\ and\  \citenamefont
  {Rigetti}}]{1908.08054}%
  \BibitemOpen
  \bibfield  {author} {\bibinfo {author} {\bibfnamefont {Keri~A.}\ \bibnamefont
  {McKiernan}}, \bibinfo {author} {\bibfnamefont {Erik}\ \bibnamefont {Davis}},
  \bibinfo {author} {\bibfnamefont {M.~Sohaib}\ \bibnamefont {Alam}}, \ and\
  \bibinfo {author} {\bibfnamefont {Chad}\ \bibnamefont {Rigetti}},\
  }\href@noop {} {\enquote {\bibinfo {title} {Automated quantum programming via
  reinforcement learning for combinatorial optimization},}\ } (\bibinfo {year}
  {2019}),\ \Eprint {http://arxiv.org/abs/arXiv:1908.08054} {arXiv:1908.08054}
  \BibitemShut {NoStop}%
\bibitem [{ \citenamefont {Khairy}\ \emph {et~al.}(2019) \citenamefont {Khairy},
   \citenamefont {Shaydulin},  \citenamefont {Cincio},  \citenamefont {Alexeev},\
  and\  \citenamefont {Balaprakash}}]{1911.04574}%
  \BibitemOpen
  \bibfield  {author} {\bibinfo {author} {\bibfnamefont {Sami}\ \bibnamefont
  {Khairy}}, \bibinfo {author} {\bibfnamefont {Ruslan}\ \bibnamefont
  {Shaydulin}}, \bibinfo {author} {\bibfnamefont {Lukasz}\ \bibnamefont
  {Cincio}}, \bibinfo {author} {\bibfnamefont {Yuri}\ \bibnamefont {Alexeev}},
  \ and\ \bibinfo {author} {\bibfnamefont {Prasanna}\ \bibnamefont
  {Balaprakash}},\ }\href@noop {} {\enquote {\bibinfo {title}
  {Reinforcement-learning-based variational quantum circuits optimization for
  combinatorial problems},}\ } (\bibinfo {year} {2019}),\ \Eprint
  {http://arxiv.org/abs/arXiv:1911.04574} {arXiv:1911.04574} \BibitemShut
  {NoStop}%
\bibitem [{ \citenamefont {Lin}\ \emph {et~al.}(2018) \citenamefont {Lin},
   \citenamefont {Lai},\ and\  \citenamefont {Li}}]{1812.10797}%
  \BibitemOpen
  \bibfield  {author} {\bibinfo {author} {\bibfnamefont {Jian}\ \bibnamefont
  {Lin}}, \bibinfo {author} {\bibfnamefont {Zhong~Yuan}\ \bibnamefont {Lai}}, \
  and\ \bibinfo {author} {\bibfnamefont {Xiaopeng}\ \bibnamefont {Li}},\
  }\href@noop {} {\enquote {\bibinfo {title} {Reinforcement-learning-based
  architecture for automated quantum adiabatic algorithm design},}\ } (\bibinfo
  {year} {2018}),\ \Eprint {http://arxiv.org/abs/arXiv:1812.10797}
  {arXiv:1812.10797} \BibitemShut {NoStop}%
\bibitem [{ \citenamefont {Beloborodov}\ \emph {et~al.}(2020) \citenamefont
  {Beloborodov},  \citenamefont {Ulanov},  \citenamefont {Foerster},
   \citenamefont {Whiteson},\ and\  \citenamefont {Lvovsky}}]{hh}%
  \BibitemOpen
  \bibfield  {author} {\bibinfo {author} {\bibfnamefont {Dmitrii}\ \bibnamefont
  {Beloborodov}}, \bibinfo {author} {\bibfnamefont {Alexander~E}\ \bibnamefont
  {Ulanov}}, \bibinfo {author} {\bibfnamefont {Jakob~N}\ \bibnamefont
  {Foerster}}, \bibinfo {author} {\bibfnamefont {Shimon}\ \bibnamefont
  {Whiteson}}, \ and\ \bibinfo {author} {\bibfnamefont {AI}~\bibnamefont
  {Lvovsky}},\ }\bibfield  {title} {\enquote {\bibinfo {title} {Reinforcement
  learning enhanced quantum-inspired algorithm for combinatorial
  optimization},}\ }\href@noop {} {\bibfield  {journal} {\bibinfo  {journal}
  {arXiv preprint arXiv:2002.04676}\ } (\bibinfo {year} {2020})}\BibitemShut
  {NoStop}%
\bibitem [{ \citenamefont {Ayanzadeh}\ \emph {et~al.}(2020) \citenamefont
  {Ayanzadeh},  \citenamefont {Halem},\ and\  \citenamefont
  {Finin}}]{ayanzadeh2020reinforcement}%
  \BibitemOpen
  \bibfield  {author} {\bibinfo {author} {\bibfnamefont {Ramin}\ \bibnamefont
  {Ayanzadeh}}, \bibinfo {author} {\bibfnamefont {Milton}\ \bibnamefont
  {Halem}}, \ and\ \bibinfo {author} {\bibfnamefont {Tim}\ \bibnamefont
  {Finin}},\ }\href@noop {} {\enquote {\bibinfo {title} {Reinforcement quantum
  annealing: A quantum-assisted learning automata approach},}\ } (\bibinfo
  {year} {2020}),\ \Eprint {http://arxiv.org/abs/arXiv:2001.00234}
  {arXiv:2001.00234} \BibitemShut {NoStop}%
\bibitem [{ \citenamefont {Sutton}\ and\  \citenamefont
  {Barto}(2018)}]{10.5555/3312046}%
  \BibitemOpen
  \bibfield  {author} {\bibinfo {author} {\bibfnamefont {Richard~S.}\
  \bibnamefont {Sutton}}\ and\ \bibinfo {author} {\bibfnamefont {Andrew~G.}\
  \bibnamefont {Barto}},\ }\href@noop {} {\emph {\bibinfo {title}
  {Reinforcement Learning: An Introduction}}}\ (\bibinfo  {publisher} {A
  Bradford Book},\ \bibinfo {address} {Cambridge, MA, USA},\ \bibinfo {year}
  {2018})\BibitemShut {NoStop}%
\bibitem [{ \citenamefont {Kaelbling}\ \emph {et~al.}(1996) \citenamefont
  {Kaelbling},  \citenamefont {Littman},\ and\  \citenamefont
  {Moore}}]{Kaelbling1996}%
  \BibitemOpen
  \bibfield  {author} {\bibinfo {author} {\bibfnamefont {L.~P.}\ \bibnamefont
  {Kaelbling}}, \bibinfo {author} {\bibfnamefont {M.~L.}\ \bibnamefont
  {Littman}}, \ and\ \bibinfo {author} {\bibfnamefont {A.~W.}\ \bibnamefont
  {Moore}},\ }\bibfield  {title} {\enquote {\bibinfo {title} {Reinforcement
  learning: A survey},}\ }\href {\doibase 10.1613/jair.301} {\bibfield
  {journal} {\bibinfo  {journal} {J. Artif. Intell. Res.}\ }\textbf {\bibinfo
  {volume} {4}},\ \bibinfo {pages} {237--285} (\bibinfo {year}
  {1996})}\BibitemShut {NoStop}%
\bibitem [{ \citenamefont {van Otterlo}\ and\  \citenamefont
  {Wiering}(2012)}]{vanOtterlo2012}%
  \BibitemOpen
  \bibfield  {author} {\bibinfo {author} {\bibfnamefont {Martijn}\ \bibnamefont
  {van Otterlo}}\ and\ \bibinfo {author} {\bibfnamefont {Marco}\ \bibnamefont
  {Wiering}},\ }\bibfield  {title} {\enquote {\bibinfo {title} {Reinforcement
  learning and markov decision processes},}\ }in\ \href {\doibase
  10.1007/978-3-642-27645-3_1} {\emph {\bibinfo {booktitle} {Adaptation,
  Learning, and Optimization}}}\ (\bibinfo  {publisher} {Springer Berlin
  Heidelberg},\ \bibinfo {year} {2012})\ pp.\ \bibinfo {pages}
  {3--42}\BibitemShut {NoStop}%
\bibitem [{ \citenamefont {Mnih}\ \emph {et~al.}(2013) \citenamefont {Mnih},
   \citenamefont {Kavukcuoglu},  \citenamefont {Silver},  \citenamefont {Graves},
   \citenamefont {Antonoglou},  \citenamefont {Wierstra},\ and\  \citenamefont
  {Riedmiller}}]{1312.5602}%
  \BibitemOpen
  \bibfield  {author} {\bibinfo {author} {\bibfnamefont {Volodymyr}\
  \bibnamefont {Mnih}}, \bibinfo {author} {\bibfnamefont {Koray}\ \bibnamefont
  {Kavukcuoglu}}, \bibinfo {author} {\bibfnamefont {David}\ \bibnamefont
  {Silver}}, \bibinfo {author} {\bibfnamefont {Alex}\ \bibnamefont {Graves}},
  \bibinfo {author} {\bibfnamefont {Ioannis}\ \bibnamefont {Antonoglou}},
  \bibinfo {author} {\bibfnamefont {Daan}\ \bibnamefont {Wierstra}}, \ and\
  \bibinfo {author} {\bibfnamefont {Martin}\ \bibnamefont {Riedmiller}},\
  }\href@noop {} {\enquote {\bibinfo {title} {Playing atari with deep
  reinforcement learning},}\ } (\bibinfo {year} {2013}),\ \Eprint
  {http://arxiv.org/abs/arXiv:1312.5602} {arXiv:1312.5602} \BibitemShut
  {NoStop}%
\bibitem [{ \citenamefont {Mnih}\ \emph {et~al.}(2015) \citenamefont {Mnih},
   \citenamefont {Kavukcuoglu},  \citenamefont {Silver},  \citenamefont {Rusu},
   \citenamefont {Veness},  \citenamefont {Bellemare},  \citenamefont {Graves},
   \citenamefont {Riedmiller},  \citenamefont {Fidjeland},  \citenamefont
  {Ostrovski},  \citenamefont {Petersen},  \citenamefont {Beattie},  \citenamefont
  {Sadik},  \citenamefont {Antonoglou},  \citenamefont {King},  \citenamefont
  {Kumaran},  \citenamefont {Wierstra},  \citenamefont {Legg},\ and\
   \citenamefont {Hassabis}}]{Mnih2015}%
  \BibitemOpen
  \bibfield  {author} {\bibinfo {author} {\bibfnamefont {Volodymyr}\
  \bibnamefont {Mnih}}, \bibinfo {author} {\bibfnamefont {Koray}\ \bibnamefont
  {Kavukcuoglu}}, \bibinfo {author} {\bibfnamefont {David}\ \bibnamefont
  {Silver}}, \bibinfo {author} {\bibfnamefont {Andrei~A.}\ \bibnamefont
  {Rusu}}, \bibinfo {author} {\bibfnamefont {Joel}\ \bibnamefont {Veness}},
  \bibinfo {author} {\bibfnamefont {Marc~G.}\ \bibnamefont {Bellemare}},
  \bibinfo {author} {\bibfnamefont {Alex}\ \bibnamefont {Graves}}, \bibinfo
  {author} {\bibfnamefont {Martin}\ \bibnamefont {Riedmiller}}, \bibinfo
  {author} {\bibfnamefont {Andreas~K.}\ \bibnamefont {Fidjeland}}, \bibinfo
  {author} {\bibfnamefont {Georg}\ \bibnamefont {Ostrovski}}, \bibinfo {author}
  {\bibfnamefont {Stig}\ \bibnamefont {Petersen}}, \bibinfo {author}
  {\bibfnamefont {Charles}\ \bibnamefont {Beattie}}, \bibinfo {author}
  {\bibfnamefont {Amir}\ \bibnamefont {Sadik}}, \bibinfo {author}
  {\bibfnamefont {Ioannis}\ \bibnamefont {Antonoglou}}, \bibinfo {author}
  {\bibfnamefont {Helen}\ \bibnamefont {King}}, \bibinfo {author}
  {\bibfnamefont {Dharshan}\ \bibnamefont {Kumaran}}, \bibinfo {author}
  {\bibfnamefont {Daan}\ \bibnamefont {Wierstra}}, \bibinfo {author}
  {\bibfnamefont {Shane}\ \bibnamefont {Legg}}, \ and\ \bibinfo {author}
  {\bibfnamefont {Demis}\ \bibnamefont {Hassabis}},\ }\bibfield  {title}
  {\enquote {\bibinfo {title} {Human-level control through deep reinforcement
  learning},}\ }\href {\doibase 10.1038/nature14236} {\bibfield  {journal}
  {\bibinfo  {journal} {Nature}\ }\textbf {\bibinfo {volume} {518}},\ \bibinfo
  {pages} {529--533} (\bibinfo {year} {2015})}\BibitemShut {NoStop}%
\bibitem [{ \citenamefont {Schulman}\ \emph {et~al.}(2017) \citenamefont
  {Schulman},  \citenamefont {Wolski},  \citenamefont {Dhariwal},  \citenamefont
  {Radford},\ and\  \citenamefont {Klimov}}]{schulman}%
  \BibitemOpen
  \bibfield  {author} {\bibinfo {author} {\bibfnamefont {John}\ \bibnamefont
  {Schulman}}, \bibinfo {author} {\bibfnamefont {Filip}\ \bibnamefont
  {Wolski}}, \bibinfo {author} {\bibfnamefont {Prafulla}\ \bibnamefont
  {Dhariwal}}, \bibinfo {author} {\bibfnamefont {Alec}\ \bibnamefont
  {Radford}}, \ and\ \bibinfo {author} {\bibfnamefont {Oleg}\ \bibnamefont
  {Klimov}},\ }\bibfield  {title} {\enquote {\bibinfo {title} {Proximal policy
  optimization algorithms},}\ }\href@noop {} {\bibfield  {journal} {\bibinfo
  {journal} {arXiv preprint arXiv:1707.06347}\ } (\bibinfo {year}
  {2017})}\BibitemShut {NoStop}%
\bibitem [{ \citenamefont {Vodopivec}\ \emph {et~al.}(2017) \citenamefont
  {Vodopivec},  \citenamefont {Samothrakis},\ and\  \citenamefont
  {Ster}}]{Vodopivec2017}%
  \BibitemOpen
  \bibfield  {author} {\bibinfo {author} {\bibfnamefont {Tom}\ \bibnamefont
  {Vodopivec}}, \bibinfo {author} {\bibfnamefont {Spyridon}\ \bibnamefont
  {Samothrakis}}, \ and\ \bibinfo {author} {\bibfnamefont {Branko}\
  \bibnamefont {Ster}},\ }\bibfield  {title} {\enquote {\bibinfo {title} {On
  monte carlo tree search and reinforcement learning},}\ }\href {\doibase
  10.1613/jair.5507} {\bibfield  {journal} {\bibinfo  {journal} {J. Artif.
  Intell. Res.}\ }\textbf {\bibinfo {volume} {60}},\ \bibinfo {pages}
  {881--936} (\bibinfo {year} {2017})}\BibitemShut {NoStop}%
\bibitem [{ \citenamefont {Morales}\ \emph {et~al.}(2019) \citenamefont
  {Morales},  \citenamefont {Biamonte},\ and\  \citenamefont
  {Zimborás}}]{1909.03123}%
  \BibitemOpen
  \bibfield  {author} {\bibinfo {author} {\bibfnamefont {Mauro E.~S.}\
  \bibnamefont {Morales}}, \bibinfo {author} {\bibfnamefont {Jacob}\
  \bibnamefont {Biamonte}}, \ and\ \bibinfo {author} {\bibfnamefont {Zoltán}\
  \bibnamefont {Zimborás}},\ }\href@noop {} {\enquote {\bibinfo {title} {On
  the universality of the quantum approximate optimization algorithm},}\ }
  (\bibinfo {year} {2019}),\ \Eprint {http://arxiv.org/abs/arXiv:1909.03123}
  {arXiv:1909.03123} \BibitemShut {NoStop}%
\bibitem [{ \citenamefont {F{\"o}sel}\ \emph {et~al.}(2018) \citenamefont
  {F{\"o}sel},  \citenamefont {Tighineanu},  \citenamefont {Weiss},\ and\
   \citenamefont {Marquardt}}]{fosel2018reinforcement}%
  \BibitemOpen
  \bibfield  {author} {\bibinfo {author} {\bibfnamefont {Thomas}\ \bibnamefont
  {F{\"o}sel}}, \bibinfo {author} {\bibfnamefont {Petru}\ \bibnamefont
  {Tighineanu}}, \bibinfo {author} {\bibfnamefont {Talitha}\ \bibnamefont
  {Weiss}}, \ and\ \bibinfo {author} {\bibfnamefont {Florian}\ \bibnamefont
  {Marquardt}},\ }\bibfield  {title} {\enquote {\bibinfo {title} {Reinforcement
  learning with neural networks for quantum feedback},}\ }\href@noop {}
  {\bibfield  {journal} {\bibinfo  {journal} {Phys. Rev. X}\ }\textbf {\bibinfo
  {volume} {8}},\ \bibinfo {pages} {031084} (\bibinfo {year}
  {2018})}\BibitemShut {NoStop}%
\bibitem [{ \citenamefont {Nautrup}\ \emph {et~al.}(2019) \citenamefont
  {Nautrup},  \citenamefont {Delfosse},  \citenamefont {Dunjko},  \citenamefont
  {Briegel},\ and\  \citenamefont {Friis}}]{nautrup2019optimizing}%
  \BibitemOpen
  \bibfield  {author} {\bibinfo {author} {\bibfnamefont {Hendrik~Poulsen}\
  \bibnamefont {Nautrup}}, \bibinfo {author} {\bibfnamefont {Nicolas}\
  \bibnamefont {Delfosse}}, \bibinfo {author} {\bibfnamefont {Vedran}\
  \bibnamefont {Dunjko}}, \bibinfo {author} {\bibfnamefont {Hans~J}\
  \bibnamefont {Briegel}}, \ and\ \bibinfo {author} {\bibfnamefont {Nicolai}\
  \bibnamefont {Friis}},\ }\bibfield  {title} {\enquote {\bibinfo {title}
  {Optimizing quantum error correction codes with reinforcement learning},}\
  }\href@noop {} {\bibfield  {journal} {\bibinfo  {journal} {Quantum}\ }\textbf
  {\bibinfo {volume} {3}},\ \bibinfo {pages} {215} (\bibinfo {year}
  {2019})}\BibitemShut {NoStop}%
\bibitem [{ \citenamefont {Xu}\ \emph {et~al.}(2019) \citenamefont {Xu},
   \citenamefont {Li},  \citenamefont {Liu},  \citenamefont {Wang},  \citenamefont
  {Yuan},\ and\  \citenamefont {Wang}}]{xu2019generalizable}%
  \BibitemOpen
  \bibfield  {author} {\bibinfo {author} {\bibfnamefont {Han}\ \bibnamefont
  {Xu}}, \bibinfo {author} {\bibfnamefont {Junning}\ \bibnamefont {Li}},
  \bibinfo {author} {\bibfnamefont {Liqiang}\ \bibnamefont {Liu}}, \bibinfo
  {author} {\bibfnamefont {Yu}~\bibnamefont {Wang}}, \bibinfo {author}
  {\bibfnamefont {Haidong}\ \bibnamefont {Yuan}}, \ and\ \bibinfo {author}
  {\bibfnamefont {Xin}\ \bibnamefont {Wang}},\ }\bibfield  {title} {\enquote
  {\bibinfo {title} {Generalizable control for quantum parameter estimation
  through reinforcement learning},}\ }\href@noop {} {\bibfield  {journal}
  {\bibinfo  {journal} {npj Quantum Inf.}\ }\textbf {\bibinfo {volume} {5}},\
  \bibinfo {pages} {1--8} (\bibinfo {year} {2019})}\BibitemShut {NoStop}%
\bibitem [{ \citenamefont {Walln{\"o}fer}\ \emph {et~al.}(2019) \citenamefont
  {Walln{\"o}fer},  \citenamefont {Melnikov},  \citenamefont {D{\"u}r},\ and\
   \citenamefont {Briegel}}]{wallnofer2019machine}%
  \BibitemOpen
  \bibfield  {author} {\bibinfo {author} {\bibfnamefont {Julius}\ \bibnamefont
  {Walln{\"o}fer}}, \bibinfo {author} {\bibfnamefont {Alexey~A}\ \bibnamefont
  {Melnikov}}, \bibinfo {author} {\bibfnamefont {Wolfgang}\ \bibnamefont
  {D{\"u}r}}, \ and\ \bibinfo {author} {\bibfnamefont {Hans~J}\ \bibnamefont
  {Briegel}},\ }\href@noop {} {\enquote {\bibinfo {title} {Machine learning for
  long-distance quantum communication},}\ } (\bibinfo {year} {2019}),\ \Eprint
  {http://arxiv.org/abs/arXiv preprint arXiv:1904.10797} {arXiv preprint
  arXiv:1904.10797} \BibitemShut {NoStop}%
\bibitem [{ \citenamefont {Karanikolas}\ and\  \citenamefont
  {Kawabata}(2018)}]{karanikolas2018improved}%
  \BibitemOpen
  \bibfield  {author} {\bibinfo {author} {\bibfnamefont {Vasilios}\
  \bibnamefont {Karanikolas}}\ and\ \bibinfo {author} {\bibfnamefont {Shiro}\
  \bibnamefont {Kawabata}},\ }\bibfield  {title} {\enquote {\bibinfo {title}
  {Improved performance of quantum annealing by a diabatic pulse
  application},}\ }\href@noop {} {\bibfield  {journal} {\bibinfo  {journal}
  {arXiv:1806.08517}\ } (\bibinfo {year} {2018})}\BibitemShut {NoStop}%
\bibitem [{ \citenamefont {King}\ \emph {et~al.}(2019) \citenamefont {King},
   \citenamefont {Yarkoni},  \citenamefont {Raymond},  \citenamefont {Ozfidan},
   \citenamefont {King},  \citenamefont {Nevisi},  \citenamefont {Hilton},\ and\
   \citenamefont {McGeoch}}]{king2019jp}%
  \BibitemOpen
  \bibfield  {author} {\bibinfo {author} {\bibfnamefont {James}\ \bibnamefont
  {King}}, \bibinfo {author} {\bibfnamefont {Sheir}\ \bibnamefont {Yarkoni}},
  \bibinfo {author} {\bibfnamefont {Jack}\ \bibnamefont {Raymond}}, \bibinfo
  {author} {\bibfnamefont {Isil}\ \bibnamefont {Ozfidan}}, \bibinfo {author}
  {\bibfnamefont {Andrew~D}\ \bibnamefont {King}}, \bibinfo {author}
  {\bibfnamefont {Mayssam~Mohammadi}\ \bibnamefont {Nevisi}}, \bibinfo {author}
  {\bibfnamefont {Jeremy~P}\ \bibnamefont {Hilton}}, \ and\ \bibinfo {author}
  {\bibfnamefont {Catherine~C}\ \bibnamefont {McGeoch}},\ }\bibfield  {title}
  {\enquote {\bibinfo {title} {Quantum annealing amid local ruggedness and
  global frustration},}\ }\href@noop {} {\bibfield  {journal} {\bibinfo
  {journal} {Journal of the Physical Society of Japan}\ }\textbf {\bibinfo
  {volume} {88}},\ \bibinfo {pages} {061007} (\bibinfo {year}
  {2019})}\BibitemShut {NoStop}%
  \bibitem{brady2021optimal}
L.~T. Brady, C.~L. Baldwin, A.~Bapat, Y.~Kharkov, and A.~V. Gorshkov, ``Optimal
  protocols in quantum annealing and quantum approximate optimization algorithm
  problems,'' \emph{Physical Review Letters}, vol. 126, no.~7, p. 070505, 2021.
  
  
  
\bibitem [{ \citenamefont {Hogg}(2003)}]{Hogg2003}%
  \BibitemOpen
  \bibfield  {author} {\bibinfo {author} {\bibfnamefont {Tad}\ \bibnamefont
  {Hogg}},\ }\bibfield  {title} {\enquote {\bibinfo {title} {Adiabatic quantum
  computing for random satisfiability problems},}\ }\href@noop {} {\bibfield
  {journal} {\bibinfo  {journal} {Phys. Rev. A}\ }\textbf {\bibinfo {volume}
  {67}},\ \bibinfo {pages} {022314} (\bibinfo {year} {2003})}\BibitemShut
  {NoStop}%
\bibitem [{ \citenamefont {{\v{Z}}nidari{\v{c}}}(2005)}]{nidari2005}%
  \BibitemOpen
  \bibfield  {author} {\bibinfo {author} {\bibfnamefont {Marko}\ \bibnamefont
  {{\v{Z}}nidari{\v{c}}}},\ }\bibfield  {title} {\enquote {\bibinfo {title}
  {Scaling of the running time of the quantum adiabatic algorithm for
  propositional satisfiability},}\ }\href@noop {} {\bibfield  {journal}
  {\bibinfo  {journal} {Phys. Rev. A}\ }\textbf {\bibinfo {volume} {71}},\
  \bibinfo {pages} {062305} (\bibinfo {year} {2005})}\BibitemShut {NoStop}%
\bibitem [{ \citenamefont {Kirkpatrick}\ and\  \citenamefont
  {Selman}(1994)}]{Kirkpatrick1994}%
  \BibitemOpen
  \bibfield  {author} {\bibinfo {author} {\bibfnamefont {S.}~\bibnamefont
  {Kirkpatrick}}\ and\ \bibinfo {author} {\bibfnamefont {B.}~\bibnamefont
  {Selman}},\ }\bibfield  {title} {\enquote {\bibinfo {title} {Critical
  behavior in the satisfiability of random boolean expressions},}\ }\href
  {\doibase 10.1126/science.264.5163.1297} {\bibfield  {journal} {\bibinfo
  {journal} {Science}\ }\textbf {\bibinfo {volume} {264}},\ \bibinfo {pages}
  {1297--1301} (\bibinfo {year} {1994})}\BibitemShut {NoStop}%
\bibitem [{ \citenamefont {Monasson}\ \emph {et~al.}(1999) \citenamefont
  {Monasson},  \citenamefont {Zecchina},  \citenamefont {Kirkpatrick},
   \citenamefont {Selman},\ and\  \citenamefont {Troyansky}}]{Monasson1999}%
  \BibitemOpen
  \bibfield  {author} {\bibinfo {author} {\bibfnamefont {R{\'{e}}mi}\
  \bibnamefont {Monasson}}, \bibinfo {author} {\bibfnamefont {Riccardo}\
  \bibnamefont {Zecchina}}, \bibinfo {author} {\bibfnamefont {Scott}\
  \bibnamefont {Kirkpatrick}}, \bibinfo {author} {\bibfnamefont {Bart}\
  \bibnamefont {Selman}}, \ and\ \bibinfo {author} {\bibfnamefont {Lidror}\
  \bibnamefont {Troyansky}},\ }\bibfield  {title} {\enquote {\bibinfo {title}
  {Determining computational complexity from characteristic `phase
  transitions'},}\ }\href {\doibase 10.1038/22055} {\bibfield  {journal}
  {\bibinfo  {journal} {Nature}\ }\textbf {\bibinfo {volume} {400}},\ \bibinfo
  {pages} {133--137} (\bibinfo {year} {1999})}\BibitemShut {NoStop}%
\bibitem [{ \citenamefont {Brockman}\ \emph {et~al.}(2016) \citenamefont
  {Brockman},  \citenamefont {Cheung},  \citenamefont {Pettersson},  \citenamefont
  {Schneider},  \citenamefont {Schulman},  \citenamefont {Tang},\ and\
   \citenamefont {Zaremba}}]{1606.01540}%
  \BibitemOpen
  \bibfield  {author} {\bibinfo {author} {\bibfnamefont {Greg}\ \bibnamefont
  {Brockman}}, \bibinfo {author} {\bibfnamefont {Vicki}\ \bibnamefont
  {Cheung}}, \bibinfo {author} {\bibfnamefont {Ludwig}\ \bibnamefont
  {Pettersson}}, \bibinfo {author} {\bibfnamefont {Jonas}\ \bibnamefont
  {Schneider}}, \bibinfo {author} {\bibfnamefont {John}\ \bibnamefont
  {Schulman}}, \bibinfo {author} {\bibfnamefont {Jie}\ \bibnamefont {Tang}}, \
  and\ \bibinfo {author} {\bibfnamefont {Wojciech}\ \bibnamefont {Zaremba}},\
  }\href@noop {} {\enquote {\bibinfo {title} {Openai gym},}\ } (\bibinfo {year}
  {2016}),\ \Eprint {http://arxiv.org/abs/arXiv:1606.01540} {arXiv:1606.01540}
  \BibitemShut {NoStop}%
\bibitem [{ \citenamefont {Dhariwal}\ \emph {et~al.}(2017) \citenamefont
  {Dhariwal},  \citenamefont {Hesse},  \citenamefont {Klimov},  \citenamefont
  {Nichol},  \citenamefont {Plappert},  \citenamefont {Radford},  \citenamefont
  {Schulman},  \citenamefont {Sidor},  \citenamefont {Wu},\ and\  \citenamefont
  {Zhokhov}}]{baselines}%
  \BibitemOpen
  \bibfield  {author} {\bibinfo {author} {\bibfnamefont {Prafulla}\
  \bibnamefont {Dhariwal}}, \bibinfo {author} {\bibfnamefont {Christopher}\
  \bibnamefont {Hesse}}, \bibinfo {author} {\bibfnamefont {Oleg}\ \bibnamefont
  {Klimov}}, \bibinfo {author} {\bibfnamefont {Alex}\ \bibnamefont {Nichol}},
  \bibinfo {author} {\bibfnamefont {Matthias}\ \bibnamefont {Plappert}},
  \bibinfo {author} {\bibfnamefont {Alec}\ \bibnamefont {Radford}}, \bibinfo
  {author} {\bibfnamefont {John}\ \bibnamefont {Schulman}}, \bibinfo {author}
  {\bibfnamefont {Szymon}\ \bibnamefont {Sidor}}, \bibinfo {author}
  {\bibfnamefont {Yuhuai}\ \bibnamefont {Wu}}, \ and\ \bibinfo {author}
  {\bibfnamefont {Peter}\ \bibnamefont {Zhokhov}},\ }\href@noop {} {\enquote
  {\bibinfo {title} {Openai baselines},}\ }\bibinfo {howpublished}
  {\url{https://github.com/openai/baselines}} (\bibinfo {year}
  {2017})\BibitemShut {NoStop}%
\bibitem [{ \citenamefont {Lloyd}(2018)}]{1812.11075}%
  \BibitemOpen
  \bibfield  {author} {\bibinfo {author} {\bibfnamefont {Seth}\ \bibnamefont
  {Lloyd}},\ }\href@noop {} {\enquote {\bibinfo {title} {Quantum approximate
  optimization is computationally universal},}\ } (\bibinfo {year} {2018}),\
  \Eprint {http://arxiv.org/abs/arXiv:1812.11075} {arXiv:1812.11075}
  \BibitemShut {NoStop}%
  \bibitem{farhi2002quantum}
E.~Farhi, J.~Goldstone, and S.~Gutmann, ``Quantum adiabatic evolution
  algorithms versus simulated annealing,'' \emph{arXiv preprint
  quant-ph/0201031}, 2002.

\bibitem{kong2017performance}
L.~Kong and E.~Crosson, ``The performance of the quantum adiabatic algorithm on
  spike hamiltonians,'' \emph{International Journal of Quantum Information},
  vol.~15, no.~02, p. 1750011, 2017.
  
  

\bibitem{roland2002quantum}
J.~Roland and N.~J. Cerf, ``Quantum search by local adiabatic evolution,''
  \emph{Physical Review A}, vol.~65, no.~4, p. 042308, 2002.

  
  \bibitem{nautrup2019optimizing}
H.~P. Nautrup, N.~Delfosse, V.~Dunjko, H.~J. Briegel, and N.~Friis,
  ``Optimizing quantum error correction codes with reinforcement learning,''
  \emph{Quantum}, vol.~3, p. 215, 2019.

\bibitem{kanno2019many}
S.~Kanno and T.~Tada, ``Many-body calculations for periodic materials via
  quantum machine learning,'' \emph{arXiv preprint arXiv:1911.10330}, 2019.
  
 \bibitem{caneva2011chopped}
T.~Caneva, T.~Calarco, and S.~Montangero, ``Chopped random-basis quantum
  optimization,'' \emph{Physical Review A}, vol.~84, no.~2, p. 022326, 2011.
  
  

  
\end{thebibliography}
%

\section*{Author contributions}

YQC wrote the codes, performed the simulations, and analyzed the data. YC wrote the codes. All authors contributed to interpreting data and engaged in useful scientific discussions. CYH conceived and supervised this project. All authors contributed to the writing.

\onecolumngrid
\section{supplementary material}

\subsection{Stochastic Descent}
We use stochastic descent (SD) algorithm to sample the AQC ground energy landscape minima with local optimal evolution schedules as the benchmark of the results acquired using MCTS. 
SD is a simple algorithm for schedule optimization in discretized search spaces. The algorithm start from a  randomly generated initial schedule and perform local field update consistently.   The neighbor schedule is accepted $\mathbf{x}\rightarrow \mathbf{x'}$ if it is better than the current one: \(\left\langle\psi(T)^\mathbf{x'}\left|H_{\text {final}}\right| \psi(T)^\mathbf{x'}\right\rangle<\left\langle\psi(T)^\mathbf{x}\left|H_{\text {final}}\right| \psi(T)^\mathbf{x}\right\rangle\).  The algorithm stops after a given number of iterations or when there is no better solution when looking up all the neighbors. The obtained schedule is a local minimum respect to local updates. In the main text we perform SD multiples times with different initial random schedules for the same search space.

\subsection{AQC evolution schedule design by Reinforcement learning}
Reinforcement Learning (RL) has been a very useful approach for the automation of complex tasks in recent years.  We briefly introduce all RL algorithms, benchmarked against Monte Carlo Tree Search and newly proposed Quantum Zero algorithms discussed in this work. Noting that all RL algorithms can be formulated as a Markov Decision Process (MDP), we first introduce the common framework before discussing specific details of each RL algorithm. 

For the automated design of annealing schedules, the MDP is given by ,
\begin{enumerate}
\item Observable space \(S\). Environment state \(s_t \in S\)  at a given timestep \(t\) with a duration $t \in [0,T]$. 
Here \(s_t=\textbf{x}\) is a vector of path parameters.
\item Action space \(A\) from which an agent picks an action \(a_t \in A\) and applies it to a state to get \(s_{t+1}\)  at each time step t. 
Action space here is  \(\{-l,-l+\Delta, l-\Delta, l\}\), where \(\pm l\) are the upper and lower bound set for the amplitudes of each frequency component with \(\Delta\) as the discretized interval.
\item A scalar reward of 1 for the annealed energy satisfying  \(|E(T)-E_{target}|\ll\epsilon\) with $\epsilon$ a given parameter, and -1 otherwise. 
$E(T)$ is the environment's feedback, taken as the expectation value of $\bra{\psi_\mathbf{x}(T)} H_{final} \ket{\psi_\mathbf{x}(T)}$.
\end{enumerate}

\textbf{Deep Q-Network (DQN)} DQN \cite{1312.5602,Mnih2015} algorithm combines RL with a deep neural network to learn a complex state-action relation in order to accomplish complex tasks. DQN was the first RL algorithm to demonstrate superhuman performance in an Atari game.  DQN overcomes unstable learning for nonlinear function approximators such as neural networks by using two techniques: experience replay and target network. Experience replay stores past experiences including state transitions, rewards and actions. These experiences are organized in mini-batches when training neural networks. The mini-batches reduce correlations between experiences used in updating deep neural networks. Target-network technique fixes parameters of a target function and replaces them with the latest network at regular intervals.
The DQN network takes states \(\{s_t\}\) as an input, and outputs a Q-value for each action, the target Q-value:
\begin{equation}\label{eq:DQN}
\hat{Q}_{k}\left(s_{t}, a_{t}\right) \leftarrow r_{t+1}+\gamma \max _{\alpha \in A} \hat{Q}_{k-1}\left(s_{t+1}, a, \theta\right)
\end{equation}
The goal for a DQN agent is to maximize expectation of its perceived reward by learning from past examples and formulate an optimal policy. We use the OpenAI Baselines  \cite{1606.01540,baselines} to train  DQN agents to design the annealing path following a Markov Decision Process. In the subsection E of Result (in the main text),  we choose discount factor \(\gamma=0.99\),  neural network of two dense layers $\{64,64\}$, learning rate \(lr=0.001\), final value of random action probability  0.01 for the DQN model.

\textbf{Advantage Actor-Critic (A2C)} Actor-Critics  \cite{10.5555/3312046} aim to take advantage of all the good stuff from both value-based RL and policy-based RL by efficiently learns an approximation for both action policy and value functions.
A2C framework contains two networks. 
One of them (actor network) is to produce the best action for a given state. The second network (critic network) learns the advantage value of taking an action as shown in Eq:\ref{eq:A2C}:
\begin{equation} \label{eq:A2C}
A\left(s_{t}, a_{t}\right) \leftarrow r_{t+1}+\gamma V_{v}\left(s_{t+1}\right)-V_{v}\left(s_{t}\right)
\end{equation}
The goal of an A2C agent is also to maximize the expectation of its perceived reward by learning from known examples. We use the OpenAI Baselines \cite{1606.01540,baselines} to train A2C agents to design annealing paths. In the subsection E of Result (in the main text), we choose discount factor \(\gamma=0.99\), neural network of two dense layer $\{64,64\}$, and a learning rate \(lr=0.001\).

\textbf{Proximal policy optimization (PPO)} 
For policy-based RL, when using gradient descent to optimize a policy objective function, the policy is usually hard to be properly updated leading to gradients vanishing or exploding. PPO  \cite{schulman}  tries to compute an update that ensuring the deviation from the previous policy relatively mild. It makes updated policy lying within a trust region and avoids additional overhead to the optimization problem by incorporating a constraint inside the objective function as a penalty. In the PPO  framework, the inaccuracy brought by occasional violations of the constraints is generally mild, 
and the computation is much simpler.
We use the OpenAI Baselines \cite{1606.01540,baselines} to train PPO agents to design the annealing paths. In the subsection E of Result (in the main text), we choose discount factor \(\gamma=0.99\), neural network of two dense layer $\{64,64\}$.

\subsection{Analysis on the transferability of optimal pulses across 3-SAT problems } \label{app: }
\begin{figure}[htp]
\centering
\includegraphics[width=0.85\textwidth,height=0.35\textwidth]{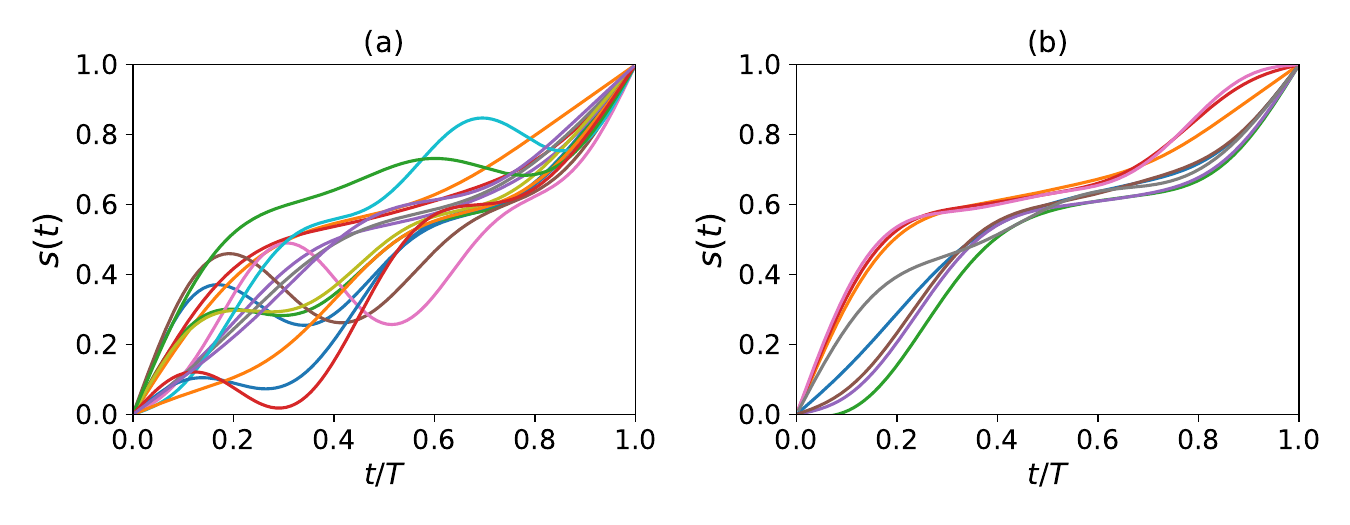}
\caption{ (a) Optimal schedules for different 3-SAT instances ($m/n=3$) in training dataset at $T=40$. (b) Top-8 average  optimal schedules for 3-SAT instances ($m/n=3$) in training dataset at $T=40$.  }
\label{fig:path}
\end{figure}

In the main text, we discuss the transferability of 'optimal' pulses across 3-SAT problems. When $T=100$ is large, the transferability is good as adiabatic evolution is likely to dominate when the annealing schedule progress slowly enough (and any paths seem to work well). However, as the annealing time budget is reduced, such as $T=40$ in the main text, directly trasnfer either an optimal schedule from a random instance or an average optimal schedule from a training dataset does not perform as well. In this non-adiabatic regime, Qzero algorithm trained with the proper transfer learning manifests its superiority. Clearly, different 3-SAT instances (even with the same $m/n=3$ ratio) possess their own uniqueness such that a direct transfer of an optimal path from another similar problem may not work when the annealing time is not sufficiently long. This point is further analyzed in Fig. \ref{fig:path}(a), in which we present some of the optimal schedules for different 3-SAT instances used for transfer learning in the main text.  In Fig. \ref{fig:path}(b), we show the top-8  average optimal schedules of the training instances. Those average optimal schedules are 'smoother' than the single optimal ones, and they successfully capture some common features across many instances in order to attain a better transferability on average. For instance, all considered 3-SAT problems tend to exhibit minimal gaps cluster around the same point (recalled dimensionless time unit) along the path as shown in Fig.7(c) in the main text. Nevertheless, it is clear that these smooth paths usually deviate from the single optimal ones as shown in Fig. \ref{fig:path}(a).


\subsection{Analysis on the Learning efficiency with system size } \label{app: }

\begin{figure}[htp]
\centering
\includegraphics[width=0.85\textwidth,height=0.35\textwidth]{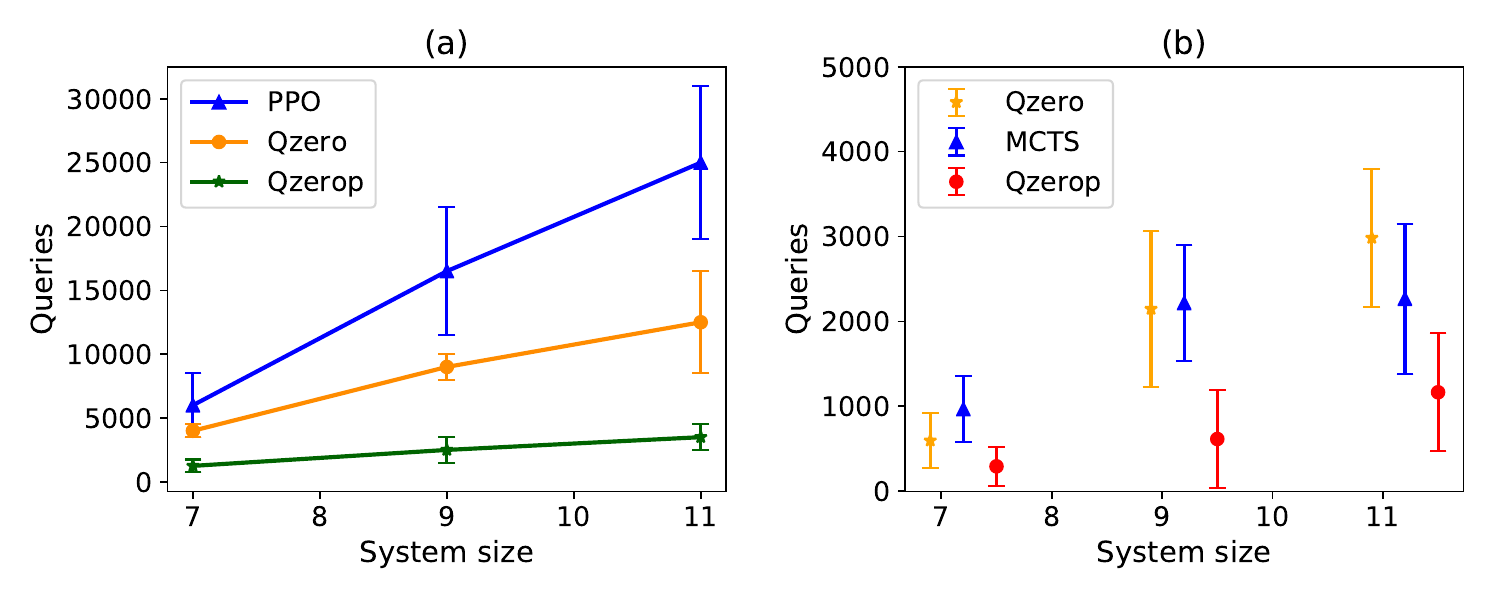}
\caption{(a) Convergence efficiencies (queries) of Qzero (Qzero-nopre), Qzerop (Qzero-pre) and PPO, on 3-SAT problems with system size $n=7, 9, 11$ under evolution time $T=70, 100, 300$ respectively.  We test four  different 3-SAT instances of $m/n=3$  for each system size and each algorithms is repeatedly learned for 10 times. (b)  Learning efficiencies  (queries) of MCTS, Qzero (Qzero-nopre), Qzerop (Qzero-pre) needed first-solution-time that satisfying $F>0.99, F>0.99, F>0.97$ for 3-SAT problems with system size $n=7$ (under $T=70$),$n=9$ (under $T=100$),$n=11$ (under $T=300$) respectively. Each algorithm is repeatedly learned for 30 times.}
\label{fig:size}
\end{figure}

In the main text we compare the learning efficiencies of MCTS, Qzero-nopre, Qzero-pre and some other common RL algorithms like PPO for 3-SAT problems with the system size $n=7$. More precisely, we examine the convergence efficiencies for training these different RL algoirthms. In order to study the scaling behaviors of those algorithms, here we present the convergence efficiencies of 3-SAT problems with system size $n=7, 9, 11$ in Fig.~\ref{fig:size}(a) . Since, according to the results presented in the main text, PPO performs best among the other tested RL algorithms, here we only compare Qzero-nopre, Qzero-pre with PPO. As the system size grows, the resource consumption for the learning increases for all algorithms, but Qzero and Qzerop increasingly outperform PPO.   

Next, we also investigate the search efficency between MCTS and Qzero-pre (enhanced with the transfer learning). For this analysis, we count the number of queries required for the first-solution-time. In other words, we stop the algorithms as soon as they find a solution that certain criterion. More precisely, in Fig.~\ref{fig:size}(b), the first-solution-time is defined as the first time reaching a solution that satisfying $F>0.99, F>0.99, F>0.97$ for 3-SAT problems with system size $n=7$ (under $T=70$),$n=9$ (under $T=100$),$n=11$ (under $T=300$), respectively. As shown, when the system size scales up, Qzerop  consistently outperform pure MCTS.  


\subsection{Analysis on the learning efficiency with the size for the state space} \label{app: }

In the main text, when running the simulation experiments in the frequency domain, we always truncate the frequency at $M=5$. Here we add more frequency components to investigate how the algorithms behave under the new simualtion setting. As the problem does not really change, merely scaling up the state space only make sense if we raise the bar for a success optimization with a higher fidelity. Since Qzero is built upon MCTS, here we only compare the learning efficiency of MCTS and PPO with respect to the scaling of the size for the state space.
\begin{figure}[htp]
\centering
\includegraphics[width=0.55\textwidth,height=0.35\textwidth]{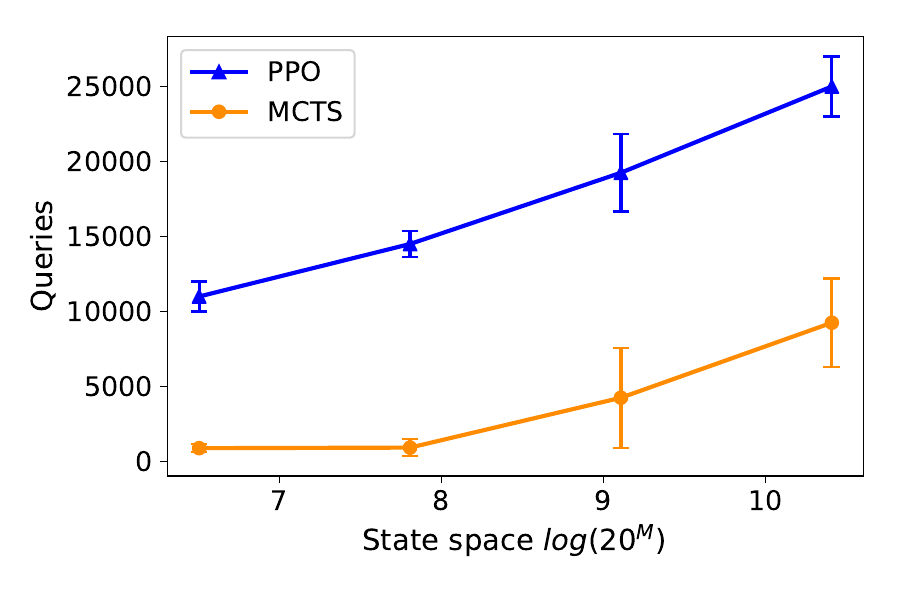}
\caption{Number of queries needed by  MCTS and PPO to converge to fidelity of 0.98 different state spaces with  frequency truncation $M=5,6,7,8$, thus state spaces $20^{5}, 20^{6}, 20^{7}, 20^{8}$. }
\label{fig:space}
\end{figure}

As shown in Fig.~\ref{fig:space}, MCTS consistently outperform PPO as the state size scales up from $20^{5}$ to $ 20^{8}$.

\subsection{Inspecting the MCTS solutions in closer details: 
Grover's search } \label{app:  }

 \begin{figure}[htp]
\centering
\includegraphics[width=0.55\textwidth,height=0.35\textwidth]{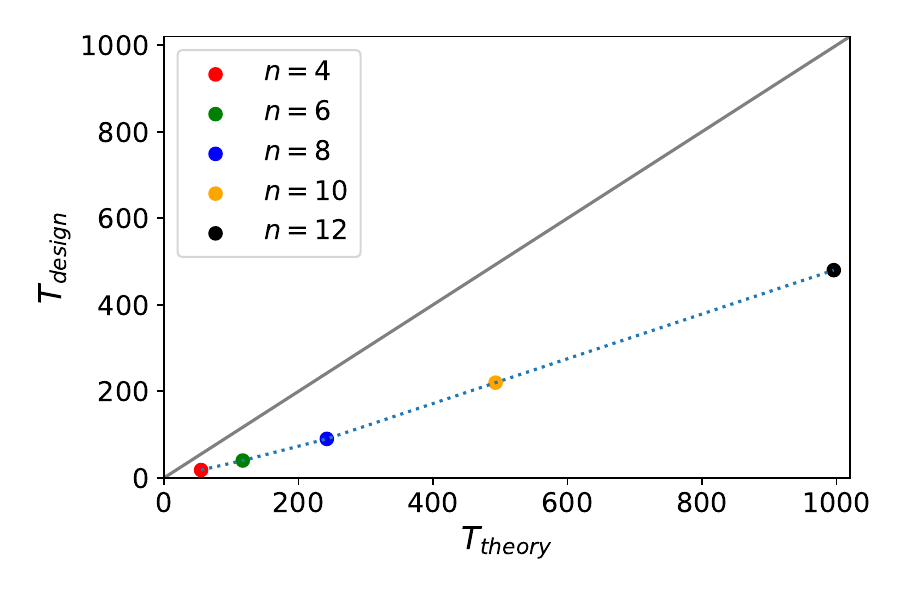}
\caption{The evolution time needed by theoretical non-linear path and MCTS-based RL designed evolution schedules for system size of $n=4,6,8,10,12$. }
\label{fig:time}
\end{figure}

\begin{figure}[htp]
\centering
\includegraphics[width=0.7\textwidth,height=0.5\textwidth]{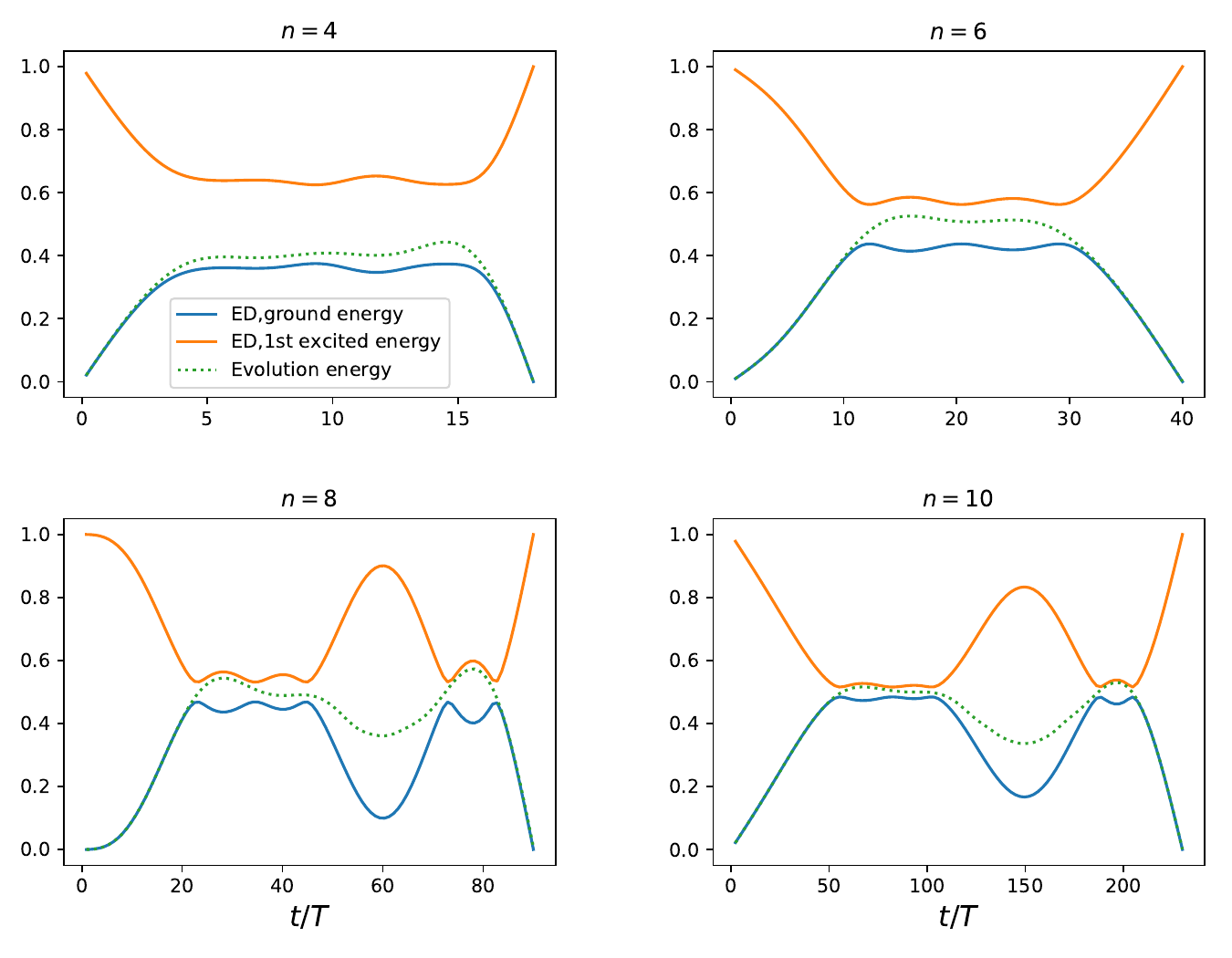}
\caption{The energy spectrum of instantaneous Hamiltonian during the evolution process for $n=4,6,8,10$.}
\label{fig:energy}
\end{figure}
For some problem, such as the example of Grover's search, there exists analytically derived non-linear annealing schedules that obey adiabaticiy. It will be interesting to analyze how paths proposed by MCTS differs from these theoretically motivated path designs. 

The problem Hamiltonian of Grover's problem for the adiabatic quantum computation is $H_{final}=\mathbb{I}-\left| b\right> \left< b \right|$, where $\left| b\right >$ is a product state in Pauli-$Z$ basis that encodes the search target. The  initial trivial Hamiltonian $H_{init}=\mathbb{I}-\left | \psi _{0}\right> \left< \psi_{0} \right|$, where $\left |\psi_{0}\right>$ is the product in Pauli-$X$ basis with all $n$ eigenvalues that equals to $1$.  Given in Ref. \cite{roland2002quantum}, the theoretically proposed non-linear path reads
\begin{equation} \label{eq: }
s(t)=\frac{1}{2}+\frac{1}{2 \sqrt{N-1}} \tan \left [  2t \epsilon \sqrt{N-1} /N  +\arctan \sqrt{N-1} \right ],
\end{equation}
where $N=2^{n}$ is the dimension of corresponding Hilbert space, $\epsilon^2$ denotes the final infidelity $\epsilon^{2}=1-F$. For the theoretically derived path, it is known that the minimally reuqired evolution time for a given $\epsilon$ scales as, 
\begin{equation} \label{eq: }
T_{theory}=\frac{1}{\varepsilon} \frac{N}{\sqrt{N-1}} \arctan \sqrt{N-1}.
 \end{equation}
 
As we can see from Fig.~\ref{fig:time}, to  realize  the same fidelity $F=0.99$, the annealing schedules designed by MCTS entails much shorter evolution time compared with the time needed by theoretically proposed non-linear path.

We further present the exact diagonalization of energy spectrum for the instantaneous Hamiltonian $H=(1-s(t))H_{init}+s(t)H_{final}$ along the adiabatic path, as well as the evolution of the  expectation value of the instantaneous Hamiltonian with respect to the instantaneous state $\left<\psi_{t} \right | H \left | \psi_{t} \right >$ in Fig.~\ref{fig:energy} for various system size of $n=4,6,8,10$ for the evolution time $T=18,50,90,200$ respectively. As clearl revealed in Fig.~\ref{fig:energy},  the accelerated paths (designed by MCTS) clearly invoke non-adiabatic transitions with the expected energies lying above the instantenous ground-state energies.

\subsection{Quantum annealing simulation in noisy environment } \label{app: }
In practice, the adiabatic device usually works in a noisy environment. Here we study the influence of  noise on the finally acquired simulation fidelity by adding noises on the rewards during the training of the algorithms.
\begin{figure}[htp]
\centering
\includegraphics[width=0.5\textwidth,height=0.35\textwidth]{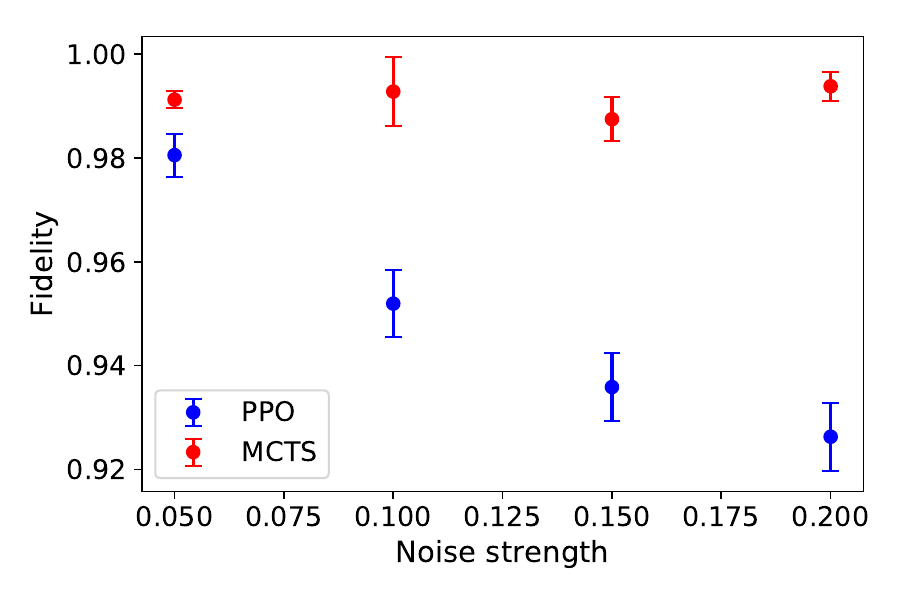}
\caption{The finally fidelity acquired by algorithms change along with noise strength.}
\label{fig:noise}
\end{figure}

In Fig.~\ref{fig:noise}, we present the the final fidelities acquired by MCST and PPO under the influence of weak-to-moderate noise levels for 3-SAT problem of system size $n=7$ and evolution time $T=70$. As shown in the figure, the path designed by MCTS manifests better noise resilience than that by PPO.


\subsection{Performing MCTS in the Quantum Circuit Model} \label{app: circuit }

In the main text, we discuss a quantum annealer operating as an analogue device. When a quantum annealing process following a given schedule \(s(t)\), it can be understood as follows.  The system is initialized in the ground state of a simple Hamiltonian, and let it evolve for a total annealing time \(T\) under the action of \(H(s(t))=(1-s(t)) H_{i n i t}+s (t)H_{f i n a l}\). The corresponding evolution operator:
$$
U(t, 0)=\mathcal{T} \exp \left(-\frac{i}{\hbar} \int_{0}^{t} \mathrm{d} t^{\prime} H\left(s\left(t^{\prime}\right)\right)\right)
$$
where \(\mathcal{T}\)exp denotes  the time-ordered exponential. Quantum annealing with a smooth schedule can be easily discretized into a digital version. \(s(t)\) can be approximated with \(K\) values \(s_1,\dots,s_K\) corresponding to evolution times \(\Delta t_1,\dots,\Delta t_K \) , with \( s_j \in (0,1]\) and \(\sum_{j=1}^{K} \Delta t_{j}=T\). The evolution operator \(U(T, 0)\) then reads
\begin{equation}
U(T, 0) \Longrightarrow U_{\mathrm{step}}=\prod_{j=1}^{\leftarrow \mathrm{K}} \mathrm{e}^{-\frac{i}{\hbar} H\left(s_{j}\right) \Delta t_{j}}
\end{equation}
where the arrow \(\leftarrow\)denotes a time-ordered product. A further digitalization step is to perform a Trotter splitting of the term \(\mathrm{e}^{-\frac{i}{h} H\left(s_{j}\right) \Delta t_{j}}\). For instance, the lowest-order Trotter splitting
\begin{equation}
\mathrm{e}^{-\frac{i}{\hbar} H\left(s_{j}\right) \Delta t_{j}} \simeq \mathrm{e}^{-i \beta_{j} H_{init}} \mathrm{e}^{-i \gamma_{j} H_{final}}+O\left(\left(\Delta t_{j}\right)^{2}\right)
\end{equation}
with \(\gamma_{j}=s_{j} \frac{\Delta t_{j}}{\hbar},\beta_{j}=\left(1-s_{j}\right) \frac{\Delta t_{j}}{\hbar}\) leads to an approximated evolution operator of the form,
\begin{equation}\label{eq:digitizedannealing}
U(T, 0) \approx U_{\mathrm{digit}}(\boldsymbol{\gamma}, \boldsymbol{\beta})=U\left(\gamma_{\mathrm{K}}, \beta_{\mathrm{K}}\right) \cdots U\left(\gamma_{1}, \beta_{1}\right)
\end{equation}
with \(U\left(\gamma_{j}, \beta_{j}\right) \equiv U_{j}=\mathrm{e}^{-i \beta_{j} H_{init}} \mathrm{e}^{-i \gamma_{j} H_{final}}\). The parameters satisfy \(\sum_{j=1}^{\mathrm{K}}\left(\gamma_{j}+\beta_{j}\right)=\frac{T}{\hbar}\).  Hence, one can easily use MCTS to design annealing schedule then perform corresponding unitary transformation to a quantum circuit.

Next, we note the proposed MCTS method can be adapted to the popular Quantum Approximate Optimization Algorithm (QAOA) method \cite{1411.4028,1812.11075,1909.03123}. QAOA is a hybrid quantum-classical algorithm that combines stae prepation in quantum circuits with classical optimization of the circuit parameters to solve the kind of combinatorial optimizations considered in this work.  A QAOA circuit (with depth P) alternates the application of  \(H_{init}\) and \(H_{final}\) to prepare a variational quantum state for P times,
\begin{equation}
\left|\psi_{\mathrm{P}}(\boldsymbol{\gamma}, \boldsymbol{\beta})\right\rangle=U\left(\gamma_{\mathrm{P}}, \beta_{\mathrm{P}}\right) \cdots U\left(\gamma_{1}, \beta_{1}\right)\left|\psi_{0}\right\rangle,
\end{equation}
where $\ket{\psi_{0}}$ is a chosen initialization. Obviously, the QAOA circuit is analogous to a digitized  quantum  annealing. When depth $P$ is sufficiently deep, the parameter \(\boldsymbol{\gamma}, \boldsymbol{\beta}\)  of  QAOA circuit can be directly taken from a correspondingly digitized  quantum annealing process, as shown in Eq:\ref{eq:digitizedannealing}; otherwise, these parameters should be obtained by optimizing the cost function  \(E_{\mathrm{P}}(\boldsymbol{\gamma}, \boldsymbol{\beta})=\left\langle\psi_{\mathrm{P}}(\boldsymbol{\gamma}, \boldsymbol{\beta})\left|H_{final}\right| \psi_{\mathrm{P}}(\boldsymbol{\gamma}, \boldsymbol{\beta})\right\rangle\) with respect to the parameters.
By drawing the analogy between QAOA (with sufficiently long P-depth) and digitized  quantum  annealing, the  proposed  MCTS  approaches may be easily applied to suggest QAOA parameters for initialization. When $P$-depth is shallow, one may even directly discretize the search space for $(\boldsymbol{\gamma},\boldsymbol{\beta})$ and perform MCTS on it.

\subsection{MCTS-designed annealing schedules in the time-frequency domain} \label{app: bang-anneal-bang}
In this section, we study the annealing schedules designed by MCTS in the time-frequency domain. 
Here we use ”Hamming weight with a spike” as an example to illustrate how one may design annealing schedule in the time-frequency domain.
”Hamming weight with a spike” is commonly used in the comparisons of quantum and classical heuristic optimization algorithms. An insight gained from this problem is that many classical search algorithms often stuck in a local minimum and fail to discover the global minimum while a quantum algorithm can.  \cite{farhi2002quantum, kong2017performance}. 

The ”Hamming weight with a spike” Hamiltonian reads:
\begin{equation}
H_{final }=\sum_{\mathbf{z} \in\{0,1\}^{n}} c(w)|\mathbf{z}\rangle\langle\mathbf{z}|,
\end{equation}
where $w=\left|\mathbf{z}\right|=\left|z_{1} \ldots z_{n}\right|$ is the  Hamming weight of a $n$-bit string. $c(w)$ is the potential 
given by  a ramp $r(w)=w$, plus a rectangular “spike” function $s(w)$ centered at $w=n/4$, for two exponents $\alpha, \beta \in [0,1]$ 
\begin{equation}
\begin{array}{l}\text { Ramp: } r(w)=w, \text { Spike: } s(w)=\left\{\begin{array}{l}n^{\beta}, \text { if } w \in\left[\frac{n}{4}-\frac{n^{\alpha}}{2}, \frac{n}{4}+\frac{n^{\alpha}}{2}\right] \\ 0, \text { otherwise. }\end{array}\right. \\ \text { Full Potential: } c(w)=r(w)+s(w)\end{array}
\end{equation}

 In this illustration, we impose $s_0(t)$ to take on a linear schedule for the most part but switch to a bang-bang control for a brief interval at the beginning and the end of the schedule as outlined below,
\begin{equation}\mathrm{s}(\mathrm{t})=
\left\{
             \begin{array}{lr}
             1 &  0\leq t \textless t_{1}, T-t_{3}-t_{4}\leq t \textless T-t_{4},\\
             0 & t_{1}\leq t \textless t_{1}+t_{2}, T-t_{4}\leq t \textless T,\\
             \frac{t}{T}+\sum_{i=1}^{M} x_{i} \sin \frac{i \pi t}{T} & t_{1}+t_{2}\leq t \textless T-t_{3}-t_{4}.
             \end{array}
\right.
\end{equation}
The optimal control problem is now expanded to assigning values to the sequence \(\mathbf{x}=\{t_{1},t_{2},t_{3},t_{4},x_1,x_2,x_3....x_M\}\), where  $x_1,x_2,x_3....x_M$ is the strength of Fourier components  and $t_{1},t_{2},t_{3},t_{4}$ indicates the distribution of initial/final bang-bang sequences. It is obvious that when $t_{i}=0, i=1,2,3,4$ the evolution schedule is just annealing schedules in the frequency domain.

\begin{figure}[htp]
\centering
\includegraphics[width=0.9\textwidth,height=0.4\textwidth]{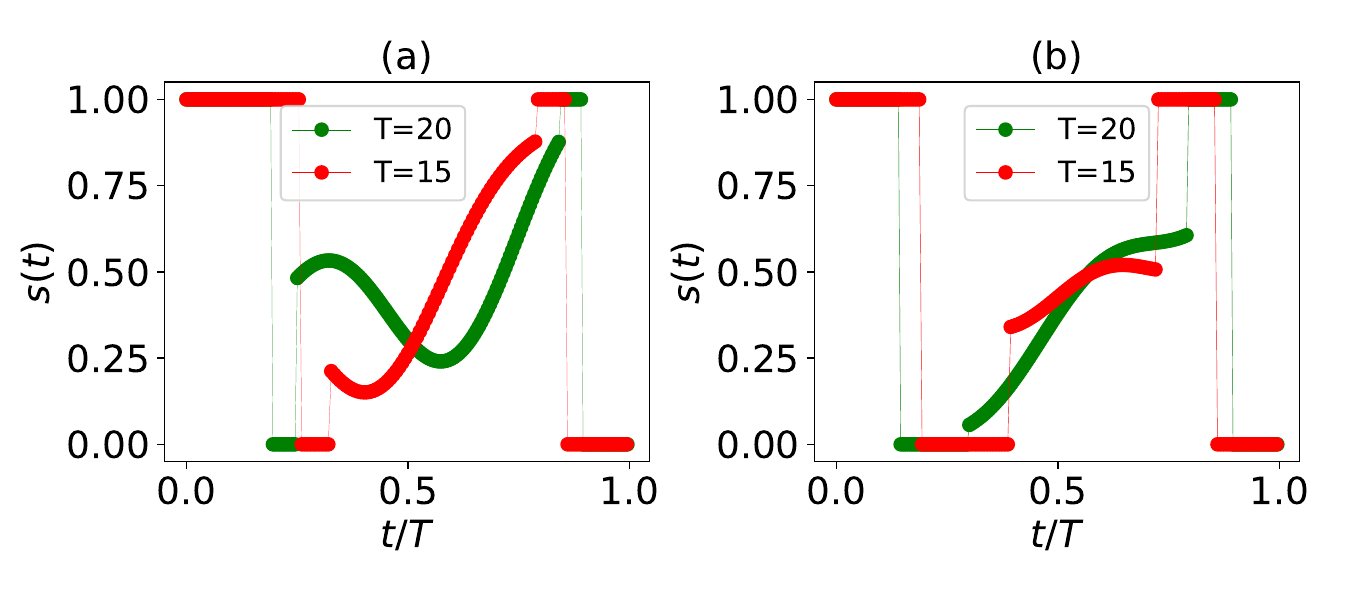}
\caption{Optimal evolution schedules designed by MCTS. (a) Results of spike instance with $n=8$, $\alpha=0.3$,  $\beta=0.2$ for $T=20,10$ respectively.  (b) Results of spike instance with $n=8$, $\alpha=0.3$,  $\beta=0.4$ for $T=20,10$ respectively. }
\label{fig:bang}
\end{figure}

Results of MCTS-designed annealing schedules for two Hamming ramp with spike problems are displayed in Fig.\ref{fig:bang}. Details regarding these Hamiltonians may be found in the caption of the figure. In the first case, Fig.\ref{fig:bang}(a), the optimal schedules  operated under a total time of $T=20$ and $T=15$ reach a final fidelity of 0.99 and 0.986 in comparison to the fidelity of 0.91 and 0.908
for a linear path, respectively.  In the second instance, shown in  Fig.\ref{fig:bang}(b), the optimal schedules, operated under a total time of $T=20 (T=15)$, reaches a final fidelity of 0.989 and 0.982, which also exceed the fidelity of 0.867 and 0.822 for a linear path, respectively.
In all these cases, MCTS recommends schedules with bang-anneal-bang profiles instead of a smooth trajectory.

\twocolumngrid

\end{document}